\documentclass[10pt,journal,compsoc]{IEEEtran}
\usepackage{tipa}
\usepackage{dsfont}
\usepackage{bbm}
\usepackage{color}
\usepackage{mathrsfs}
\usepackage{amsfonts}
\usepackage{cite,url,subfigure,epsfig,graphicx}
\usepackage{amssymb,amsmath,bm,makecell}
\usepackage{indentfirst}
\usepackage{algorithmic}
\usepackage{algorithm}
\usepackage{epstopdf}
\usepackage{tabulary}
\usepackage{booktabs}
\IEEEoverridecommandlockouts
\usepackage{times,verbatim,amsfonts,amsmath,color}
\usepackage{amsthm}

\newtheorem{lemma}{\textbf{Lemma}}
\newtheorem{theorem}{\textbf{Theorem}}

\newtheorem{assumption}{\textbf{Assumption}}
 \usepackage{multirow}
\usepackage{soul}
\DeclareMathOperator{\sign}{sgn} 
\newcommand{\norm}[1]{\left\lVert#1\right\rVert}

\begin{document}

\title{Energy Efficient Federated Learning over Heterogeneous Mobile Devices via Joint Design of Weight Quantization and Wireless Transmission}
 
\author{Rui~Chen,~\IEEEmembership{Student Member,~IEEE}
        Liang~Li,~\IEEEmembership{Member,~IEEE}
        Kaiping Xue,~\IEEEmembership{Senior Member,~IEEE} Chi Zhang,~\IEEEmembership{Member,~IEEE}
        Miao Pan,~\IEEEmembership{Senior Member,~IEEE}
        and~Yuguang Fang,~\IEEEmembership{Fellow,~IEEE}
\IEEEcompsocitemizethanks{\IEEEcompsocthanksitem R. Chen and M. Pan are with the Department of Electrical and Computer Engineering, University of Houston, Houston, TX, 77204 (e-mail: rchen19@uh.edu, mpan2@uh.edu).
\IEEEcompsocthanksitem L. Li is with the School of Computer Science, Beijing University of Posts and Telecommunications, Beijing 100876, P. R. China (e-mail: liliang1127@bupt.edu.cn).
\IEEEcompsocthanksitem K. Xue is with the School of Cyber Security, and C. Zhang is with the School of Information Science and Technology, University of Science and Technology of China, China, Hefei, 230027, P. R. China (e-mail: kpxue@ustc.edu.cn, chizhang@ustc.edu.cn).
\IEEEcompsocthanksitem Y. Fang is with the Department of Electrical and Computer Engineering, University of Florida, Gainesville, Florida 32611 (e-mail: fang@ece.ufl.edu).
}

}

\IEEEtitleabstractindextext{%
\begin{abstract}
Federated learning (FL) is a popular collaborative distributed machine learning paradigm across mobile devices. However, practical FL over resource constrained mobile devices confronts multiple challenges, e.g., the local on-device training and model updates in FL are power hungry and radio resource intensive for mobile devices. To address these challenges, in this paper, we attempt to take FL into the design of future wireless networks and develop a novel joint design of wireless transmission and weight quantization for energy efficient FL over mobile devices. Specifically, we develop flexible weight quantization schemes to facilitate on-device local training over heterogeneous mobile devices. Based on the observation that the energy consumption of local computing is comparable to that of model updates, we formulate the energy efficient FL problem into a mixed-integer programming problem where the quantization and spectrum resource allocation strategies are jointly determined for heterogeneous mobile devices to minimize the overall FL energy consumption (computation + transmissions) while guaranteeing model performance and training latency. Since the optimization variables of the problem are strongly coupled, an efficient iterative algorithm is proposed, where the bandwidth allocation and weight quantization levels are derived. Extensive simulations are conducted to verify the effectiveness of the proposed scheme.
\end{abstract}

}

\maketitle

\IEEEdisplaynontitleabstractindextext
 
\IEEEpeerreviewmaketitle

\IEEEraisesectionheading{ \section{Introduction}\label{sec:introduction}}

\IEEEPARstart{D}{ue} to the incredible surge of mobile data and the growing computing capabilities of mobile devices, it becomes a trend to apply deep learning (DL) on these devices to support fast responsive and customized intelligent applications. Recently, federated learning (FL) has been regarded as a promising DL solution to providing an efficient, flexible, and privacy-preserving learning framework over a large number of mobile devices. Under the FL framework \cite{mcmahan2017communication}, each mobile device executes model training locally and then transmits the model updates, instead of raw data, to an FL server. The server will then aggregate the intermediate results and broadcast the updated model to the participating devices. Its potential has prompted wide applications in various domains such as keyboard predictions \cite{hard2018federated}, physical hazards detection in smart home \cite{yu2020learning}, health event detection~\cite{brisimi2018federated}, etc. 
Unfortunately, it also faces many significant challenges when deploying FL over mobile devices in practice. First, although mobile devices are gradually equipped with artificial intelligence (AI) computing capabilities, the limited resources (e.g., battery power, computing and storage capacity) restrain them from training deep and complicated learning models at scale. Second, it is unclear how to establish an effective wireless network architecture to support FL over mobile devices. Finally, the power-hungry local computing and wireless communications during iterations in FL may be too much for the power-constrained mobile devices to afford.

The mismatch between the computing and storage requirements of DL models and the limited resources of mobile devices becomes even more challenging due to the increasing complexity of the state-of-art DL models. To address this issue, one of the most popular solutions is to compress a trained network \cite{li2019additive, han2015deep, gupta2015deep}. Han et al. \cite{han2015deep} successfully applied multiple compression methods, e.g., pruning and quantization, to several large-scale neural networks (e.g.,  AlexNet and VGG-16). These compression techniques help reduce model complexity by multiple orders of magnitude and speed up model inference on mobile devices. However, on-device training is less explored and more complicated than its inference counterpart. Some pioneering works \cite{zhang2018lq, fu2020don} have made efforts on quantizing the model parameters to make it possible to conduct computationally efficient on-device training. Nevertheless, most existing compressed on-device learning frameworks and the associated convergence analysis for the potential on-device training only consider the case of a single mobile device. A few works, such as \cite{hou2018analysis}, have considered quantized on-device training in distributed learning settings. However, they assign the same quantization strategy for different mobile devices. In practice, FL may encompass massively distributed mobile devices that are highly heterogeneous in computing capability and communication conditions. Thus, it is in dire need to develop a flexible quantization scheme catering to the heterogeneous devices and investigate the impacts of such heterogeneity on learning performance.

Besides the on-device training for local computing, the energy consumption for FL over mobile devices also includes the wireless communications for the intermediate model update exchanges. Particularly, with the advance of computing hardware and future wireless communication techniques, like 5G and beyond (5G+) \cite{NextG20}, we have observed that the energy consumption for local computing in FL is comparable to that for the wireless transmissions on mobile devices. For instance, the energy consumption of local computing (e.g., 42.75J for one Tesla P100 GPU of one training iteration for Alexnet with batch size of 128) is comparable to that of today's wireless communications (e.g., 38.4J for transmitting 240MB Alexnet model parameters at 100 Mbps data rate~\cite{3gpp.21.915}). Thus, a viable design of the energy efficient FL over mobile devices has to consider the energy consumption of both ``working" (i.e., local computing) and ``talking" (i.e., wireless communications). However, most existing works in wireless communities have mainly conducted the radio resource allocation under the FL convergence constraints \cite{tran2019federated,vu2020cell,chen2019joint}, while neglecting the energy consumption in learning. Moreover, among the previous works, the targeted learning models are either relatively simple (i.e., with convex loss functions) or shallow networks \cite{tran2019federated,vu2020cell,chen2019joint, yang2020federated}, which is inconsistent with the current trend of the overparameterized DL models. On the other hand, most efforts in the machine learning communities have focused on communication efficient FL algorithmic designs, such as compressing the size of the model updates or reducing the update frequency during the training phase. The basic assumption is that the wireless transmission data rate is slow, which results in the bottleneck to support complicated learning models over mobile devices. Therefore, the goal of such designs is to reduce the number of communications in model updates without considering the advance of wireless transmissions.


Fortunately, the future wireless transmissions (e.g., 5G/6G cellular, WiFi-6 or future version of WiFi), featured by very high data rate (1 Gbps or more \cite{NextG20}) with ultra low latency of 1 ms or less for massive number of devices, 
can be leveraged to relieve the communication bottleneck with proper design. Furthermore, the multi-access edge computing in the future networks enhances the computing capabilities at the edge networks, and hence provides an ideal architecture to support viable FL. 

Motivated by the aforementioned challenges (i.e., inefficient on-device training and large overall energy consumption in FL training), in this paper, we develop a wireless transmission and on-device weight quantization co-design for energy efficient FL over heterogeneous mobile devices. We aim to 1) facilitate efficient on-device training on heterogeneous local devices via a flexible quantization scheme, and 2) minimize the overall energy consumption during the FL learning process by considering the learning performance and training latency. Based on the derived convergence analysis, we formulate the energy minimization problem to determine the optimal quantization strategy and bandwidth allocations. Our major contributions are summarized as follows.
\begin{itemize}
\item We propose a novel efficient FL scheme over mobile devices to reduce the overall energy consumption in communication and computing. Briefly, subject to their current computing capacities, the participating mobile devices are allowed to compress the model and compute the gradients of the compressed version of the models. Meanwhile, for a given training time threshold, the network resource allocation is to minimize the total computing and communication energy cost in FL training. 

\item To facilitate on-device training for FL over  heterogeneous mobile devices, weight quantization is employed to best utilize the limited computing capacities by representing model parameters with different bit-widths. We further provide the theoretical analysis of the convergence rate of FL with quantization and obtain a closed-form expression for the novel convergence bound in order to explore the relationship between the weight quantization error, and the performance of the FL algorithm. 

\item Based on the obtained theoretical convergence bound, the energy minimization problem in FL training is formulated as a mixed-integer nonlinear problem to balance the computing and communication costs by jointly determining the bandwidth allocation and weight quantization levels for each mobile device. An efficient iterative algorithm is proposed with low complexity, in which we derive new closed-form solutions for the bandwidth allocation and weight quantization levels.
%
\item We evaluate the performance of our proposed solution via extensive simulations using various open datasets and models to verify the effectiveness of our proposed scheme. Compared with different schemes, our proposed method shows significant superiority in terms of energy efficiency for FL over heterogeneous devices.
\end{itemize}

The rest of this paper is organized as follows. The related work is discussed in Section \ref{sec:ReWork}. In Section~\ref{Sec:SysM}, a detailed description of the system model is presented and the convergence analysis of the proposed FL with weight quantization is also discussed. The energy minimization problem and joint quantization selection and bandwidth allocation algorithm are presented in Section~\ref{Sec:Solution}. In Section \ref{Sec:Result}, the feasible solutions from the real datasets are analyzed. The paper is concluded in Section \ref{Sec:con}.

\section{Related Work}\label{sec:ReWork}
\subsection{Cost-efficient design for FL over wireless networks}
Recognizing that training large-scale FL models over mobile devices can be both time and energy consuming tasks, several research efforts have been made on decreasing these costs via device scheduling~\cite{nishio2019client}, network optimization~\cite{yang2020federated} and resource utilization optimization~\cite{vu2020cell, shi2021towards, shi2020joint, zeng2021energy, yang2020energy, mo2021energy, li2021talk}. In particular, the resource allocation for optimizing overall FL energy efficiency was studied in~\cite{zeng2021energy, yang2020energy, mo2021energy, li2021talk}. Mo et al. in~\cite{mo2021energy} have designed the computing and communication resources allocation to minimize the energy consumption while only considering the CPU models for mobile devices. Zeng et al.~\cite{zeng2021energy} proposed to partition the computing workload between CPU-GPU to improve the computing energy efficiency. However, their resource allocation strategies are for particular (non-optimal) model parameters (i.e., weight quantization levels in this paper). Thus, they overlook the opportunities to first reduce the costs in learning (i.e., model quantization in this paper) before utilizing the available resources. Close to our work, Li et al.~\cite{li2021talk} considered to sparsify the model size before transmission to improve communication efficiency and determine heterogeneity-aware gradient sparsification strategies. However, they neglect the mismatch between the computing/storage requirement for on-device training and the limited computing resources on mobile devices. Based on the example illustrated in Section~\ref{sec:introduction}, on-device computing consumes more energy than model update transmission. Hence, different from~\cite{li2021talk}, this paper leverages the quantization method for on-device training instead of wireless transmission only.

\subsection{On-device training with low precision}
Various works have been developed for on-device learning to reduce the model complexities via low precision operation and storage requirements~\cite{de2018high}. In the extreme case, the weights and activations are represented in one bit, called Binary Neural Networks (BNN) \cite{courbariaux2016binarized}, while the performance degrades significantly in large DNNs. For weight quantization, the prior work such as “LQ-Net” in Zhang et al. \cite{zhang2018lq} quantized weights and activations such that the inner products can be computed efficiently with bit-wise operations, performing in the case of single machine computation. Similar to our work, Fu et al. \cite{fu2020don} considered the weight quantization for local devices in the distributed learning setting and proposed to quantize activations via estimating Weibull distributions. However, they did not consider optimization for energy efficiency in FL training. Besides, they assigned the same quantization level across different participating devices, which limited the performance when facing the challenges of device heterogeneity. It left the impact of flexible quantization schemes on the learning model accuracy as an open problem, which will be addressed in this work. Unlike these existing works, a mobile-compatible FL algorithm with flexible weight quantization is introduced in our proposed model. By jointly considering the heterogeneous computing and communication conditions, we formulate the overall FL energy (computing + transmissions) minimization problem to seek for the optimal weight quantization levels and bandwidth allocation across multiple mobile devices.

\section{FL with Flexible Weight Quantization}\label{Sec:SysM}
\begin{figure} \centering
  {\includegraphics[width=0.95\linewidth]{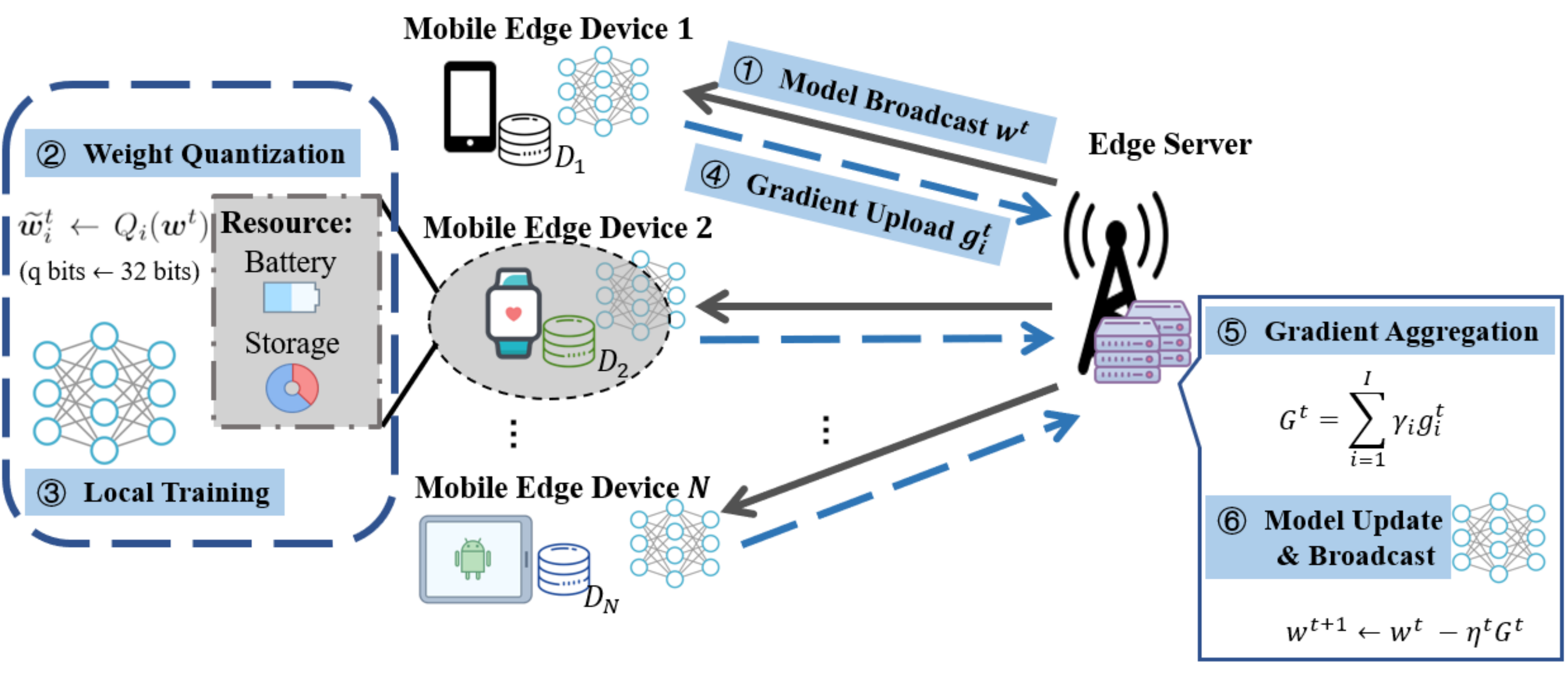}}
 \caption{Federated learning framework with weight quantization.} \label{fig:System}
\end{figure}
\subsection{Preliminary of Weight Quantization}
In this subsection, we introduce the related concepts about weight quantization for on-device training. Quantization is an attractive solution to implementing FL models on mobile devices efficiently. It represents model parameters, including the weights, feature maps, and even gradients, with low-precision arithmetic (e.g., 8-bit fixed-point numbers). When the model parameters are stored and computed with low-bitwidth, the computational units and memory storage to perform the operations during on-device training are much smaller than the full-precision counterparts, resulting in energy reduction during on-device training.

To train the FL model in low precision, we define a quantization function $Q(\cdot)$ to convert a real number $w$ into a quantized version $\hat{w}  = Q(w)$. We use the same notation for quantizing vectors since $Q$ acts on each dimension of the vector independently in the same manner. Moreover, we employ stochastic rounding (SR)~\cite{gupta2015deep} in our proposed model and analyze its convergence properties. SR, also known as unbiased rounding, possesses the important property: $\mathbb{E}[Q(w)]=w$. This property avoids the negative effect of quantization noise, which is useful for the theory of non-convex setting~\cite{li2019dimension}. For each component $w_n$ of a vector $\mathbf{w}$, the function $Q(\cdot)$ converts the data type from 32-bit into $q$-bit, defined as:
\begin{equation}
    Q(w_n) = s \cdot \sign(w_n) \cdot \left\{\begin{matrix}
                I_{a+1}, &  w.p. \quad \frac{|w_n|}{s \Delta_q}- \frac{I_a}{\Delta_q}\\ 
                I_a, & w.p. \quad \frac{I_{a+1}}{\Delta_q}-\frac{|w_n|}{s \Delta_q}
         \end{matrix}\right.,
\end{equation}
where $\sign(\cdot)$ represents the sign function, $s= \norm{\mathbf{w}}_{\infty}$ denotes the scaling factor, the index $k$ satisfies $I_a \leq \frac{|w_n|}{s} \leq I_{a+1}$, quantization set $\mathcal{S}_w = \{-I_A, \cdots,I_0,\cdots, I_A \}$ with $A = 2^{q-1}-1, 0= I_0 \leq I_1 \leq \cdots \leq I_A$ are uniformly spaced, and $\Delta_q$ denotes the quantization resolution as $\Delta_q=I_{a+1} - I_a = 1/(2^q-1)$. Smaller resolution leads to a smaller gap and keeps as much information as the original weight, while it has higher memory requirements. In practice, the bit-width for the weight quantization can be extremely small, like 2 or 3 bits without notable performance degradation. Other parameters, such as the weight gradient calculations and updates, are applied to capture accumulated small changes in stochastic gradient descent (SGD). In contrast, quantization makes them insensitive to such information and may impede convergence performance during training. Therefore, we keep a higher precision for the gradients than the weights and inputs so that the edge server aggregates the local gradients and updates the global model in full precision.

\subsection{FL with flexible weight quantization}
We consider a mobile edge network consisting of one edge server and a set $\mathcal{N} = \{1,2,\cdots,N\}$ of distributed mobile devices, collaboratively training a DNN model through FL framework, which is depicted in Fig.~\ref{fig:System}. 
Each mobile device $i$ is equipped with a single antenna and has its own dataset $\mathcal{D}_i$ with data size $|\mathcal{D}_i|$. The data is collected locally by the mobile device $i$ itself. Generally, each learning model has a particular loss function $f_j(\mathbf{w})$ with the parameter vector $\mathbf{w}$ for each data sample $j$. The loss function represents the difference of the model prediction and groundtruth of the training data. Thus, the loss function on the local data of mobile devices $i$ is given as $F_i(\mathbf{w}):= \frac{1}{|\mathcal{D}_i|} \sum_{j=1}^{|\mathcal{D}_i|} f_j(\mathbf{w}).$ The training objective of the shared model is to collaboratively learn from all the participating mobile devices, formulated as follows: 
\begin{equation}
    \min_{\mathbf{w} \in \mathbb{R}^d} F(\mathbf{w}) = \sum_{i=1}^N \pi_i F_i(\mathbf{w}), \label{FL_obj}
\end{equation}
where $d$ denotes the total number of the DNN model parameters and $\pi_i$ is the weight of the $n$-th device such that  $ \pi_i = |\mathcal{D}_i| / \sum_{i=1}^N |\mathcal{D}_i| $ and $\sum_{i=1}^N \pi_i =1$.
Given the sensitive nature of the users' data, each mobile device keeps its data locally instead of uploading its data to the edge server. An FL framework~\cite{mcmahan2017communication} is adopted to solve problem (\ref{FL_obj}), named FedAvg, that allows the users to update the model to the edge server periodically. Let $r$ be the $r$-th training iteration in FL. In FedAvg, the edge server first broadcasts the latest model $\bar{\mathbf{w}}^{r}$ to all the devices. Second, every device $i \in \mathcal{N}$ performs $H$ mini-batch SGD steps in parallel, obtains and transmits its intermediate local model ${\mathbf{w}}^{r+H}_i$ to the edge server. After that, the edge server will update the model based on aggregated results from the mobile devices, i.e., $\bar{\mathbf{w}}^{r+H} = \sum \pi_i {\mathbf{w}}^{r+H}_i$. This procedure repeats until FL converges.

Targeting at the energy-efficient FL training over mobile devices, we propose a flexible weight quantization (FWQ) scheme for heterogeneous mobile devices. After mobile devices receive the shard model from the edge server, they first quantize and store the model to satisfy their current storage budget. Unlike the prior works that maintain the same quantization strategy across all the participating devices, FWQ considers device heterogeneity and allows the mobile devices to perform weight quantization with different bit-widths of $q_i$ during on-device training and transmit the model updates in more bits. Note that the weights and gradients at the server side remain in full precision operations to avoid further model performance degradation. A pseudo-code of our FWQ algorithm is presented in Alg.~\ref{alg:OverallFramwork}.
 
\begin{algorithm}[!t] 
\caption{Flexible Weight Quantizated FL (FWQ-FL)} 
\label{alg:OverallFramwork} 
\begin{algorithmic}[1]  
\REQUIRE $\eta=$ learning rate; $Q(\cdot)=$ quantization function; initial $\bar{\mathbf{w}}^0$; a mini-batch size $M$; a number of local SGD iterations $H$; a number of training iterations $R$\\
\ENSURE $\bar{\mathbf{w}}^{R}$
\FOR{$r = 0,\cdots,R-1$}
\STATE Edge server sends $\bar{\mathbf{w}}^r$ to the set of participating mobile devices $\mathcal{N}$
\FOR{each mobile device $i \in \mathcal{N}$ in parallel}
\STATE Sample mini-batch data set $\{\widetilde{\mathbf{x}}_m,\widetilde{\mathbf{y}}_m\}_{m=1}^M$ from $\mathcal{D}_i$
\STATE Compute the mini-batch stochastic gradient $\triangledown \widetilde{f}_i ( {\mathbf{w}}_i^{r})$ 
\STATE Update the model parameters\\
$ {\boldsymbol{w}}_i^{r+1} \leftarrow Q\left( {\boldsymbol{w}}_i^{r} - \eta_i \triangledown \widetilde{f}_i \left( {\boldsymbol{w}}_i^{r}\right)\right)$ \{store the weight in the low precision\}
\IF{$((r+1) \mod H) = 0 $} 
\STATE Send ${\boldsymbol{w}}_i^{r+1}$ to the FL server. 
\ENDIF
\ENDFOR
\STATE Edge server updates the global model $\bar{\mathbf{w}}^{r+1}$ as follows\\
\STATE $\bar{\mathbf{w}}^{r+1} \leftarrow \sum_{i=1}^N \pi_i {\boldsymbol{w}}_i^{r+1}$ \{update the weight in the high precision\}
\ENDFOR
\end{algorithmic}
\end{algorithm}

\subsection{Convergence Analysis of FL with FWQ}\label{Sec:CovAls}
Before we discuss the convergence of Alg.~\ref{alg:OverallFramwork}, we make the following assumptions on the loss function, which are commonly used for the analysis of SGD approach under the distributed/federated learning settings~\cite{li2019convergence}, \cite{yu2019parallel}.

\begin{assumption} \label{as:Lsmooth}
All the loss functions $f_j$ are differentiable and their gradients are $L$-Lipschitz continuous in the sense of $l_2$-norm: for any $x$ and $y \in \mathbb{R}^d$, $\norm{\triangledown f_j(x)- \triangledown f_j(y)}_2 \leq L\norm{x-y}_2$.
\end{assumption}
\begin{assumption}\label{as:divergence}
  Assume that $\widetilde{f}_i$ is randomly sampled from the $i$-th mobile device local loss functions. For local device $i$, its stochastic gradient is an unbiased estimator and its variance: $\mathbb{E}||{\triangledown \widetilde{f}_i(\mathbf{w}^r) - \triangledown F_i(\mathbf{w}^r) }||_2^2 \leq \tau_i^2$. Thus, the a mini-batch size $M$ of gradient variance is given as $\tau_i^2/M$ and its second moment is $\mathbb{E} \norm{ {\triangledown}\widetilde{f}_i(\boldsymbol{w}^r)}_2^2 \leq G^2$, for any $i = 1,\cdots,N$.
\end{assumption}


Assumption~\ref{as:Lsmooth} indicates that the local loss functions $F_i$ and the aggregated loss function $F$ are also $L$-smooth. The unbiasedness and bounded variance of stochastic gradients in Assumption~\ref{as:divergence} are customary for non-convex analysis of SGD. 

 

\begin{theorem} \label{tm:nomcvx_convergence}
Let the learning rate $\eta$ be $\sqrt{\frac{M}{R}}$. If Assumptions \ref{as:Lsmooth}-\ref{as:divergence} hold, the average-squared gradient after $R$ 
iterations is bounded as follow,
\begin{flalign}
     &\frac{1}{R} \sum_{t=0}^{R-1} \mathbb{E}\norm{\triangledown F(\bar{\boldsymbol{w}}^{r})}_2^2 &\nonumber\\
     & \leq \frac{4(\mathbb{E} \left[F(\bar{\boldsymbol{w}}^{0}) \right] - F^{\star})}{\sqrt{MR}} + \frac{6H L\tau}{\sqrt{MR}} + \sqrt{d}LG \sum_{i=1}^N \pi_i^2 \delta_{i},& \label{convergence_rate} \\
     & \leq \mathcal{O}(\frac{H+1}{\sqrt{MR}}) + \mathcal{O}(\sqrt{d} \sum_{i=1}^N \pi_i^2 \delta_{i}),
\end{flalign}
where $\delta_i = s\Delta_{b_i}$ is the quantization noise. $\tau = \sum_{i=1}^N  \pi_i^2 \tau_i^2$, and $F^{\star}$ is the global minimum of F. 
\begin{proof}
    Please refer to the detailed proof in Appendix~\ref{PF:T1}  in the separate supplemental file.
\end{proof}
\end{theorem}
Here, the average expected squared gradient norm characterizes the convergence rate due to the non-convex objective in modern learning models~\cite{ghadimi2013stochastic, yu2019parallel, reisizadeh2020fedpaq}. From Theorem~\ref{tm:nomcvx_convergence}, we can observe that the proposed model admits the same convergence rate as parallel SGD in the sense that both of them attain the asymptotic convergence rate $\mathcal{O}(\frac{1}{\sqrt{MR}})$. Weight quantization makes FL converge to the neighborhood of the optimal solution without affecting the convergence rate. The limit point of the iterates is related to the quantization noise $\delta_i$. If the quantization becomes more fine-grained (i.e., by increasing the number of bits), the model performance will approach the model with full precision floating point. 

\section{Optimization for Energy Efficient FWQ}
\label{Sec:Solution}
Motivated by the above discussion, the quantization levels $\{q_i\}_{i=1}^N$ and the numbers of local SGD iterations, $H$, act as critical parameters of FL training performance (i.e., model convergence rate). Besides, these strategies also greatly impact the energy consumption of mobile devices since they affect the total communication rounds and computing workload per round. In this section, we formulate the energy efficient FWQ problem (EE-FWQ) under model convergence and training delay guarantee. We develop flexible weight quantization and bandwidth allocation to balance the trade-off between computing and communication energy of mobile devices in FL training. We start with discussion on the computing and communication energy model, followed by problem formulation and solution.

\subsection{Energy Model}
\subsubsection{Computing Model}
Here, we consider the GPU computing model instead of the CPU model, for two reasons. First, CPUs cannot support relatively large and complicated model training tasks. Second, GPUs are more energy efficient than CPUs for on-device training and are increasingly integrated into today's mobile devices (e.g., Google Pixel). The GPU based training makes computing energy consumption comparable to that of communications in FL. Noted that the local computing of mobile device $i$ involves the data fetching in GPU memory modules and the arithmetic in GPU core modules, where the voltage and frequency of each module are independent and configurable: 

1)	\textit{GPU runtime power model} of mobile device $i$ is modeled as a function of the core/memory voltage/frequency~\cite{mei2017energy},
\begin{flalign}
   p_i^{cp}  =  p_i^{G0} + \zeta_i^{mem} f_i^{mem} +\zeta_i^{core} (V_i^{core})^2 f_i^{core}, \label{Eq:ComP}
\end{flalign}
where $p_i^{G0}$ is the summation of the power consumption unrelated to the GPU voltage/frequency scaling; $V_i^{core}, f_i^{core},f_i^{mem}$ denote the GPU core voltage, GPU core frequency, and GPU memory frequency, respectively; $\zeta_i^{mem}$ and $\zeta_i^{core}$ are the constant coefficients that depend on the hardware and arithmetic for one training iteration, respectively.

2) \textit{GPU execution time model} of mobile device $i$ with quantization level $q_i$ is formulated as
\begin{equation}
    T^{comp}_i(q_i)  =  t_i^0 +  \frac{c_1(q_i)\theta_i^{mem}}{f_i^{mem}}  + \frac{c_2(q_i)\theta_i^{core}}{f_i^{core}}, \label{Eq:ComTime}
\end{equation}
where $t_i^0$ represents the other component unrelated to training task; $\theta_i^{mem}$ and $\theta_i^{core}$ denote the number of cycles to access data from the memory and to compute one mini-batch size of data samples, respectively, which are measured on a platform-based experiment in this paper. Due to the weight quantization, the number of cycles for data fetching and computing are reduced with scaling $c_1({q_i})$ and $c_2({q_i})$, respectively. For simplicity, we assume that the number of cycles for data fetching and computing scales, $c_1({q_i})$ and $c_2({q_i})$, are linear functions of data bit-width $q_i$, respectively. This is reasonable since the quantization reduces the bit-widths, and the data size scales linearly to the bit representation~\cite{yang2017method}.

With the above GPU power and performance model, the local energy consumed to pass a single mini-batch SGD with quantization strategy $q_i$ of the $i$-th  mobile device is the product of the runtime power and the execution time, i.e.,
\begin{equation}
    E_i^{comp}(q_i, H) = H \cdot p_i^{cp} \cdot T_i^{cp}(q_i). \label{Eq:CompEne}
\end{equation}

\subsubsection{Communication Model}
We consider orthogonal frequency-division multiple access (OFDMA) protocol for devices to upload their local results to the edge server. 
The total channel bandwidth is bounded by $B_{\max}$ and $B_{i}$ is denoted as the bandwidth allocated to device $i$ where $B_{i}$ satisfies $\sum_{i=1}^N B_{i} \leq B_{\max}$. As a result, the achievable transmission rate (bit/s) of mobile device $i$ can be calculated as
\begin{equation}
    \gamma_{i} = B_{i} \ln\left(1+\frac{h_{i} p_i^{cm}}{N_0 }\right),\label{Eq:CommRate}
\end{equation}
where $N_0$ represents the noise power, and $p_i^{cm}$ is the transmission power.
Here, $h_{i}$ denotes the average channel gain of the mobile device $i$ during the training task of FWQ-FL.
The dimension of the gradient vector $g_i$ is fixed for a given model so that the overall data size to transmit the gradient vector is the same for all the mobile devices, which is denoted by $D_g$. Then, the communication time to transmit $D_g$ for mobile device $i$ is
\begin{equation}
    T_i^{cm}(B_{i}) = \frac{D_g}{\gamma_{i}} = \frac{D_g}{B_{i} \ln\left(1+\frac{h_{i} p_i^{cm}}{N_0}\right)}. \label{Eq:CommTime}
\end{equation}
Thus, the communication energy consumption of  mobile device $i$ can be derived as 
\begin{equation}
    E_i^{comm}(B_{i})= \frac{D_g p_i^{cm}}{B_{i} \ln\left(1+\frac{h_{i}p_i^{cm}}{N_0}\right)}. \label{Eq:CommEne}
\end{equation}

\subsection{Problem Formulation}
Considering the computing capabilities of different mobile devices vary, we formulate the optimization problem to minimize the total energy consumption during the training process as 
\begin{subequations}\label{Eq:ori1}
\begin{align}
\min_{\substack{H, K, \epsilon_q, \\ \mathbf{q},\mathbf{B}}} & \quad \sum_{i=1}^N K \left( E_i^{comm}(B_{i}) + E_i^{comp}(q_i,H) \right) \label{Eq:Objective}\\
& \text{s.t.} \quad c_3(q_i)U_i \leq C_i,  \forall i \in \mathcal{N}, \label{Constr:Capacity}\\
& \qquad  A_3 \sum_{i=1}^N \pi_i^2 \delta_i \leq \epsilon_q, \label{Constr:Qerror}\\
& \qquad \frac{A_1 H +A_2}{\sqrt{MHK}} + A_3 \sum_{i=1}^N \pi_i^2 \delta_i \leq \epsilon, \label{Constr:error}\\
& \qquad \max_i \quad K \left( H T_i^{cp}  + T_{i}^{comm} \right) \leq T_{\max}, \label{Constr:Time_1}\\
& \qquad \sum_{i=1}^N B_{i} \leq B_{\max}, \label{Constr:Comm}\\
& \qquad B_{i} > 0, q_i \in \mathcal{Q}, \forall i \in \mathcal{N},\label{Constr:Range_r} \\
& \qquad H \in \mathbb{Z}^{+}, 0 \leq \epsilon_q \leq \epsilon, \label{Constr:Range_Z}    
\end{align}
\end{subequations}
where $K$ represents the total number of communication rounds, $U_i$, and $C_i$ represent the learning model size (MB) stored in full precision and the memory capacity in mobile device $i$, respectively. $c_3(q_i)$ is the ratio of the bit-width to full precision. $\mathbf{q}=[q_1,\cdots,q_N]$ and $\mathbf{B}=[B_{1},\cdots,B_{N}]$ are the quantization and bandwidth allocation strategies of mobile devices, respectively.  Constraint (\ref{Constr:Capacity}) states the model size stored on mobile device $i$ does not exceed its storage capacity. The constraint (\ref{Constr:Qerror}) controls the average quantization error over participating devices as small as possible. The constraints in (\ref{Constr:Time_1}) ensures the entire training time can be completed within predefined deadline $T_{\max}$. In constraint (\ref{Constr:Comm}), the bandwidth allocation to the mobile devices must not exceed the channel bandwidth available to the edge server. Constraints (\ref{Constr:Range_r}) and (\ref{Constr:Range_Z}) indicate that variables take the values from a set of non-negative numbers. Bit representation set $\mathcal{Q}$ is defined as a power of 2, ranging from 8 to 32 bits, which is a standard-setting and hardware friendly \cite{torti2018embedded}. The number of communication rounds $K$ is determined by the FL model convergence. Based on the results in Theorem~\ref{tm:nomcvx_convergence}, we set upper bound to satisfy the convergence constraint as in (\ref{Constr:error}), where $A_1$, $A_2$ and $A_3$ are coefficients\footnote{These coefficients can be estimated by using a small sampling set of training experimental results.} used to approximate the big-$\mathcal{O}$ in Eqn.~(\ref{convergence_rate}). Furthermore, given the constraint (\ref{Constr:Qerror}), we can rewrite the (\ref{Constr:error}) as
\begin{equation}
\frac{A_1 H +A_2}{\sqrt{MHK}} + \epsilon_q \leq \epsilon. \label{Constr:error1}
\end{equation}
For the relaxed problem, if any feasible solution $H$, $\epsilon_q$, and $K$ satisfies constraint (\ref{Constr:error1}) with inequality, we note that the objective function is a decreasing function of $K$. Thus, for optimal $K$, the constraint (\ref{Constr:error1}) is always satisfied with equality, and we can derive $K$ from this equality as
\begin{equation}
  K(H,\epsilon_q) = \frac{(A_1H+A_2)^2}{MH(\epsilon-\epsilon_q)^2}, \label{round}
\end{equation}
From (\ref{round}), we observe that $K(H,\epsilon_q)$ is a function of $H$ that first decreases and then increases, which implies that too small and too large $H$ all lead to high communication cost and that there exists an optimal $H$. Besides, a large $\epsilon_q$, which results from aggressive quantization levels (small bit-widths), also hinders the learning efficiency since it requires more communication rounds to recover the learning accuracy. In light of this, local update and weight quantization levels should be carefully determined to minimize the overall energy consumption for FWQ-FL.

For the ease of analysis, we simplify the description of the GPU time model as a linear function of $q_i$, i.e.,
$T_i^{cp}(q_i) = c_i^2 q_i + c_i^1$. By substituting (\ref{round}) into its expression, we obtain
\begin{equation}  \label{Eq:Reobj}
\begin{aligned}
\min_{\substack{H,\epsilon_q,\\ \mathbf{q},\mathbf{B}}} & \quad \sum_{i=1}^N \frac{(A_0H+A_1)^2}{MH(\epsilon-\epsilon_q)^2} \left(\frac{p^{cm}_iD_g}{\gamma_{i}} + H p_i^{cp} (c_i^2 q_i + c_i^1 ) \right)\\ 
& \text{s.t.} \quad  (\ref{Constr:Capacity})-(\ref{Constr:Range_Z}).
\end{aligned}
\end{equation}
The relaxed problem above is a mixed-integer non-linear programming. It is intractable to deal with due to the multiplicative form of the integer variables ($H$ and $\mathbf{q}$) and continuous variables ($\epsilon_q$ and $\mathbf{B}$) in both the objective function and constraints. In the following, we develop an iterative algorithm with low complexity to seek feasible solutions.


\subsection{Iterative algorithm for EE-FWQ}
The proposed iterative algorithm divides the original problem (\ref{Eq:Reobj}) into two sub-problems: 1) Local update and quantization error optimization (for $H$ and $\epsilon_q$); 2) Joint weight quantization selection and bandwidth allocation (for $\mathbf{q}$ and $\mathbf{B}$), which can be solved in an iterative manner. For the two sub-problems, we are able to derive the closed-form solutions for local updates, bandwidth allocation and weight quantization levels. The details are presented in the following subsections.
 
\subsubsection{Local update and quantization error optimization}
To obtain the optimal strategies for FWQ, we first relax $H$ as a continuous variable for theoretical analysis, which is later rounded back to the nearest integer. Given $\overline{\mathbf{B}}$ and $\overline{\mathbf{q}}$, problem (\ref{Eq:Reobj}) is written as follows 
\begin{subequations}\label{Eq:hl}
\begin{align}
\min_{H,\epsilon_q} & \quad \frac{(A_0H+A_1)^2}{MH(\epsilon-\epsilon_q)^2}(E^{cm}(\overline{\mathbf{B}})+H E^{cp}(\overline{\mathbf{q}}) ) \label{obj:hl}\\
\text{s.t.} &  \quad  \frac{(A_0H+A_1)^2}{MH(\epsilon-\epsilon_q)^2} \leq \frac{T_{\max}}{T_i^{cm}({\overline{B}_{i}})+HT_i^{cp}(\overline{q}_i)}, \forall i \in \mathcal{N},\\
& \quad \epsilon_q \geq \epsilon_q^{\min},\\
& \quad 0 \leq \epsilon_q \leq \epsilon, H\geq 0, 
\end{align}
\end{subequations}
where $\epsilon_q^{\min} = \sum_{i=1}^N  \frac{A_3 \pi_i^2 s}{2^{\overline{q}_i}-1}$, $E^{cm}(\overline{\mathbf{B}}) = \sum_{i=1}^N E_i^{cm}({\overline{B}_{i}})$, and $E^{cp}(\overline{\mathbf{q}}) =\sum_{i=1}^N E_i^{cp}(\overline{q}_i)$. 

\begin{theorem}\label{tm:oh}
The optimal $\epsilon_q^{\star}$ in problem (\ref{Eq:Reobj}) satisfies
\begin{equation}
    \epsilon_q^{\star} = \epsilon_q^{\min}, \label{eq:eq_opt}
\end{equation}
and the optimal $H^{\star}$ is given by
\begin{subequations} \label{Eq:hl_r}
\begin{align}
\min_{H} & \quad \Psi(H) \triangleq \frac{(A_0H+A_1)^2(E^{cm}(\overline{\mathbf{B}})+H E^{cp}(\overline{\mathbf{q}}) )}{MH(\epsilon-\epsilon_q^{\min})^2} \label{ob:hl_r}\\
\text{s.t.} &  \quad  H_{\min} \leq H \leq H_{\max}, 
\end{align}
\end{subequations}
where $\rho(H_{\min}) = \rho(H_{\max}) = MN(\epsilon-\epsilon_q^{\min})^2T_{\max}$ and $\rho(H)$ is defined in (\ref{c:hl}). 
\begin{proof}
Please refer to the detailed proof in Appendix~\ref{PF:oh} in the separate supplemental file. 
\end{proof}
\end{theorem}
Noted that it can be verified that the objective function $\Psi(H)$ in (\ref{Eq:hl_r}) is convex. The optimal $H^{\star}$ can be obtain by setting the following first-order derivative into zero,
{\small
\begin{flalign}
\frac{\mathrm{d} \Psi(H)}{\mathrm{d} H} &= 2A_0^2HE^{cp}(\overline{\mathbf{q}}) +  A_0^2 E^{cm}(\overline{\mathbf{B}}) + 2A_0A_1E^{cp}(\overline{\mathbf{q}})  \nonumber\\
&\quad - \frac{A_1^2E^{cm}(\overline{\mathbf{B}})}{H^2}.\label{eq:d1_H}
\end{flalign}
}
It is a cubic equation of $H$ and can be solved analytically via Cardano formula~\cite{schlote2005bl}. Therefore, for the fixed values of $\overline{\mathbf{q}}$ and $\overline{\mathbf{B}}$, we have a unique real solution of $H$ in closed form as follows 
{\small
\begin{flalign} 
H &= \sqrt[3]{\sqrt{\frac{\alpha^3\beta}{27}+\frac{\beta^2}{4}}-\frac{\alpha^3}{27}-\frac{\beta}{2}} + \sqrt[3]{-\sqrt{\frac{\alpha^3\beta}{27} +\frac{\beta^2}{4}}-\frac{\alpha^3}{27}-\frac{\beta}{2}} \nonumber\\
&\quad +\frac{\alpha}{3}, \label{eq:H_solution}
\end{flalign}
}
with $\alpha  = \frac{A_0 E^{cm}(\overline{\mathbf{B}})  +2A_1E^{cp}(\overline{\mathbf{q}})}{2A_0 E^{cp}(\overline{\mathbf{q}})},$ and $\beta  = -\frac{A_1^2 E^{cm}(\overline{\mathbf{B}}) }{2A_0^2E^{cp}(\overline{\mathbf{q}})}$.


\subsubsection{Joint weight quantization selection and bandwidth allocation}
Given the updated $H$, $\epsilon_q$, the optimal quantization levels $\mathbf{q}^{\star}$ and the bandwidth allocation $\mathbf{B}^{\star}$ can be obtained by solving the following problem,
\begin{subequations}\label{Eq:qu}
\begin{flalign}
\underset{\mathbf{{q}},\mathbf{B}}{\min} & \quad K(H,\epsilon_q) \sum_{i=1}^N  \frac{p^{cm}_i\alpha_{i}^1}{B_{i}} + Hp_i^{cp} \cdot (c_i^2 q_i + c_i^1)&\\
\text{s.t.} &  \quad (\ref{Constr:Capacity}), (\ref{Constr:Qerror}), (\ref{Constr:Comm}), (\ref{Constr:Range_r}),\\
& \quad \frac{\alpha_{i}^1}{B_{i}} + H (c_i^2 q_i + c_i^1) \leq \frac{T_{\max}}{K(H,\epsilon_q)},  \forall i \in \mathcal{N}.  \label{constr:1}
\end{flalign}
\end{subequations}

Based on the observation of problem (\ref{Eq:qu}), it is clear that problem (\ref{Eq:qu}) is a mixed-integer non-linear problem. Besides, the integer variable $q_i$ and a fractional form of continuous variable $B_i$ are linearly coupled in constraint (\ref{constr:1}), which makes the optimization problem difficult to tackle. To address the above issues, we first introduce a new variable $\widetilde{q} = \log_2(q)$ and its finite set can be defined as $\widetilde{Q} = \{1,2,3,4,5\}$. We then relax ${\widetilde{q}_i}$ to be continuous and then round the solution. Since $\widetilde{q} = \log_2(q)$ is monotonously increasing function, we can transform an equivalent formulation as follows 
\begin{subequations}\label{Eq:qt1}
\begin{align}
\underset{\mathbf{\widetilde{q}}, \mathbf{B}}{\min} & \quad K(H,\epsilon_q) \sum_{i=1}^N \frac{p^{cm}_i\alpha_{i}^1}{B_{i}} + Hp_i^{cp} (c_i^2 2^{\widetilde{q}_i} + c_i^1) &\\
\text{s.t.} &\quad (\ref{Constr:Comm}),\\
&  \quad \phi(\widetilde{q}_1,\cdots,\widetilde{q}_N) \triangleq \sum_{i=1}^N  \frac{A_3 \pi_i^2 s}{2^{2^{\widetilde{q}_i}}-1} \leq \epsilon_q, \label{Constr:newq1}& \\ 
&  \quad c_3(2^{\widetilde{q}_i}) U_i \leq C_i,  \forall i \in \mathcal{N},\label{Constr:newq2}&\\
& \quad \frac{\alpha_{i}^1}{B_{i}} + H (c_i^2 2^{\widetilde{q}_i} + c_i^1) \leq \frac{T_{\max}}{K(H,\epsilon_q)} ,  \forall i \in \mathcal{N}, &\\
&\quad B_{i} > 0, q^{\min} \leq \widetilde{q}_i \leq q^{\max}, \forall i \in \mathcal{N}. \label{Constr:newq3}
\end{align}
\end{subequations}
For objective function in (\ref{Eq:qt1}), $\frac{K(H,\epsilon_q)\alpha_i^1}{B_i}$ and $p_i^{cp}c_i^2 2^{\widetilde{q}_i}$ are convex functions in $B_i$ and $\widetilde{q}_i$, respectively. The affine combination of convex functions preserves convexity. Similarly, we can easily verify the convexity of the constraints. 

Next, we propose an efficient iterative algorithm to reduce the computational complexity. The main idea of the proposed iterative algorithm as follows. In the $(z)$-th iteration, we first fix the bandwidth in the $(z-1)$-th iteration, denoted as $\mathbf{B}^{(z-1)}$ to solve problem (\ref{Eq:qt1}) to obtain quantization strategy $\mathbf{\widetilde{q}}$; then, with the updated $\mathbf{\widetilde{q}}^{(z)}$, we can get the optimal $\mathbf{B}^{(z)}$. In the intermediate steps, we attempt to derive some analytical solutions to reduce the computation load.

\begin{algorithm}[!t] 
\caption{The proposed iterative algorithm for (\ref{Eq:qu})} 
\label{alg:iter}
\begin{algorithmic}[1]  
\STATE {\bf Input:} Given $H$, $\epsilon_q$, two small constants, $\iota_1$, $\iota_2$, and a large positive number $\hat{\mu}$.
\STATE {\bf Output:} Optimal $2^{\widetilde{q}_i^{\star}}$, ${B}^{\star}_i$
\STATE {\bf Initialization:} $\mu^1_{LB} = \omega_{LB} = 0$; $\mu^1_{UB} = \hat{\mu}$; $\omega_{UB} = \hat{\omega}$;
\STATE Set $\mu^1 = (\mu^1_{UB}+\mu^1_{LB})/2$ and $\omega = (\omega_{UB}+\omega_{LB})/2$
\REPEAT
\STATE Set $\omega = (\omega_{UB}+\omega_{LB})/2$
\STATE Calculate ${B}_i^{(z)}$ via $(\ref{Eq:opb})$
\REPEAT
\STATE Calculate $\widetilde{q}_i^{(z)}$ via $(\ref{Eq:opq})$
\IF { $\phi(\widetilde{q}_1^{(z)},\cdots, \widetilde{q}_N^{(z)}) > \epsilon_q$ }
\STATE Set $\mu^1_{UB} = \mu^1$
\ELSE
\STATE Set $\mu^1_{LB} = \mu^1$
\ENDIF
\UNTIL{$\mu^1_{UB} - \mu^1_{LB} \leq \iota_1$}
\IF { $\sum B_i^{(z)} > B$ }
\STATE Set $\omega_{UB} = \omega$
\ELSE
\STATE Set $\omega_{LB} = \omega$
\ENDIF
\UNTIL{$\omega_{UB} - \omega_{LB} \leq \iota_2$}
\end{algorithmic}
\end{algorithm}

In  the $(z)$-th iteration, we can decompose problem (\ref{Eq:qt1}) into two convex subproblems as

\begin{subequations}\label{Eq:qu1}
\begin{flalign}
\underset{\mathbf{\widetilde{q}}^{(z)}}{\min}&  \quad R \sum_{i=1}^N p_i^{cp} (c_i^2 2^{\widetilde{q}_i^{(z)}} + c_i^1)& \label{obj:qu1}\\
\text{s.t.} & \quad (\ref{Constr:newq1}),(\ref{Constr:newq2}),(\ref{Constr:newq3}),&\\
&\frac{\alpha_{i}^1}{HB_{i}^{(z-1)}}+ (c_i^2 2^{\widetilde{q}_i^{(z)}}+ c_i^1) \leq \frac{T_{\max}}{R}, \forall i \in \mathcal{N}, &
\end{flalign}
\end{subequations}
where $R= H K(H,\epsilon_q)$ and 
\begin{subequations}\label{Eq:b1}
\begin{flalign}
&\underset{\mathbf{B}^{(z)}}{\min}  \quad K(H,\epsilon_q) \sum_{i=1}^N \frac{p^{cm}_i\alpha_{i}^1}{B_{i}^{(z)}} &\\
\text{s.t.} &\quad (\ref{Constr:Comm}),(\ref{Constr:newq3}),& 
\end{flalign}
\begin{align}
&\quad  \frac{K(H,\epsilon_q)\alpha_{i}^1}{B_{i}^{(z)}} \leq T_{\max} -RT^{cp}_i( 2^{\widetilde{q}_i^{(z)}}) ,  \forall i \in \mathcal{N}.
\end{align}
\end{subequations}

\begin{theorem} \label{tm:oq}
The optimal quantization levels $\widetilde{q}^{\star}_i$ and bandwidth allocation $B^{\star}_i$ for the $i$-th device are given by
\begin{flalign}
    &\widetilde{q}_i^{(z){\star}} = \min\{\widetilde{q}^{\max}_i, \widetilde{q}_i(\mu^{1\star})\}, \label{Eq:opq}
\end{flalign}
and 
\begin{flalign}
    B_{i}^{(z){\star}} &= \max\{B_{i,\min}^{(z)}(\widetilde{q}_i^{(z){\star}}), B_{i}^{(z)}(\omega^{\star}) \}, \label{Eq:opb}
\end{flalign}
where 
\begin{flalign}
&\widetilde{q}_i  = \log_2\left(\log_2 (\lambda_i + \sqrt{\lambda_i^2+4}) - 1 \right),\\
&\lambda_i  = \frac{\ln(2)\mu^{1\star} A_3 \pi_i^2 s^2}{R(p_i^{cp}+\mu_{i}^2(\mu^{1\star})) }, \\ 
&B_{i,\min}^{(z)}(\widetilde{q}_i^{(z){\star}})  = \frac{K(H,\epsilon_q)}{T_{\max} -RT^{cp}_i( 2^{\widetilde{q}_i^{(z){\star}}}) },\\
& B_{i}^{(z)}(\omega^{\star}) = \frac{p^{cm}_i\alpha_{i}^1(A_0H+A_1)}{\omega^{\star} \sqrt{MH}(\epsilon-\epsilon_q)},
\end{flalign}
$\mu^{1\star}$ and $\omega^{\star}$ are the optimal Lagrange multipliers to satisfy the quantization error constraint $\phi(\widetilde{q}^{(z)\star}_1, \cdots, \widetilde{q}^{(z)\star}_N) = \epsilon_q$ and bandwidth capacity constraint $\sum_{i=1}^N B_{i}^{(z){\star}} = B_{\max}$, respectively. 
\begin{proof}
Please refer to the detailed proof in Appendix~\ref{PF:oq} in the separate supplemental file. 
\end{proof}
\end{theorem}
Theorem~\ref{tm:oq} suggests that $\widetilde{q}^{\star}_i$ is determined by local computing capabilities. Specifically, small quantization levels can be allocated to devices with weaker computing capabilities for the benefit of sum computing energy reduction. Given the overall quantization error constraint, the devices with higher computing capabilities may use a higher quantization level to maintain the model accuracy.
It also indicates that the optimal bandwidth allocation $\mathbf{B} $ depends not only on the channel conditions ($h_i$) but also on the quantization levels $\widetilde{q}^{\star}_i$. Concisely, assigned bandwidth increases with the poor channel condition to avoid the straggler issues. In addition, when the devices use a higher quantization level for local training (higher computing energy), the device should be assigned more bandwidth to reduce total energy consumption. 

The algorithm that solves problem (\ref{Eq:Reobj}) is summarized in Alg~\ref{alg:GBD}. Since the optimal solution of problem (\ref{Eq:hl}) or (\ref{Eq:qu}) can be obtained in each loop, the objective value of the problem (\ref{Eq:Reobj}) keeps decreasing in the loop until the solutions of the two subproblems converge. Next, we analyze the computational complexity of Alg.~\ref{alg:GBD}. To solve the EE-FWQ problem by using Alg.~\ref{alg:GBD}, two subproblems (\ref{Eq:hl}) and (\ref{Eq:qu}) need to be solved. For the subproblem (\ref{Eq:hl}), we can obtain a unique real solution of $H$ from (\ref{eq:d1_H}) in closed form, which does not resort to any iterative solver. For the subproblem (\ref{Eq:qu}), we use Alg.~\ref{alg:iter}, where the complexity is $\mathcal{O}(N\log_2(1/\iota_1)\log_2(1/\iota_2))$ with accuracy $\iota_1$. As a result, the total complexity of Alg.~\ref{alg:GBD} is $\mathcal{O}(NL\log_2(1/\iota_1)\log_2(1/\iota_2))$ where $L$ is the number of iterations required in Alg~\ref{alg:GBD}. The complexity of Alg.~\ref{alg:GBD} is low since $\mathcal{O}(NL\log_2(1/\iota_1)\log_2(1/\iota_2))$ grows linearly with the total number of participating devices.

\begin{algorithm}[t] 
\caption{Joint design of flexible weight quantization and bandwidth allocation for EE-FWQ} 
\label{alg:GBD} 
\begin{algorithmic}[1]  
\STATE {\bf Input:} Initialize $H(0), \epsilon_q(0), q_i(0), B_i(0)$ of problem~(\ref{Eq:Reobj}) and set $l = 0$. 
\STATE {\bf Output:} $H^{\star}$, $\epsilon_q^{\star}$, $\mathbf{q}^{\star}$, $\mathbf{B}^{\star}$
\REPEAT
\STATE With given $\mathbf{q}(l)$, $\mathbf{B}(l)$, compute $\epsilon_q(l+1)$ and $H(l+1)$ via (\ref{eq:eq_opt}) and (\ref{eq:H_solution}), respectively 
\STATE With given $\epsilon_q(l+1)$ and $H(l+1)$, compute ${q}_i(l+1)$ and $b_i(l+1)$ by Alg.~\ref{alg:iter}
\UNTIL objective value (\ref{Eq:Reobj}) converges
\STATE Rounding $\hat{q}_i = \lfloor  \widetilde{q}_i^{\star} \rceil$ and $\lfloor H^{\star} \rceil$ and obtain the quantization strategy $q_i^{\star} = 2^{\hat{q}_i}$.
\end{algorithmic}
\end{algorithm}

It should be noted that the FL server is in charge of solving the optimization in (\ref{Eq:ori1}). It is piratical because the FL protocol in \cite{bonawitz2019towards} requires mobile devices to check in with the FL server first before the FL training begins. Hence, the FL server can collect the information ($c_1(q_i)$, $c_2(q_i)$, $C_i$, $p_i^{cm}$ and $h_i$) from mobile devices, determine the optimal strategies ($q_i$, $B_i$) of each device via Alg.~\ref{alg:GBD}, and inform the strategies to the participating devices. It only needs to be solved once if the network information remains unchanged. That is absolutely affordable for the FL server.


\begin{figure*}[t] \centering 
  \subfigure[Testing accuracy vs epochs.\label{fig:TvI}]
  {\includegraphics[width=4.4cm]{ 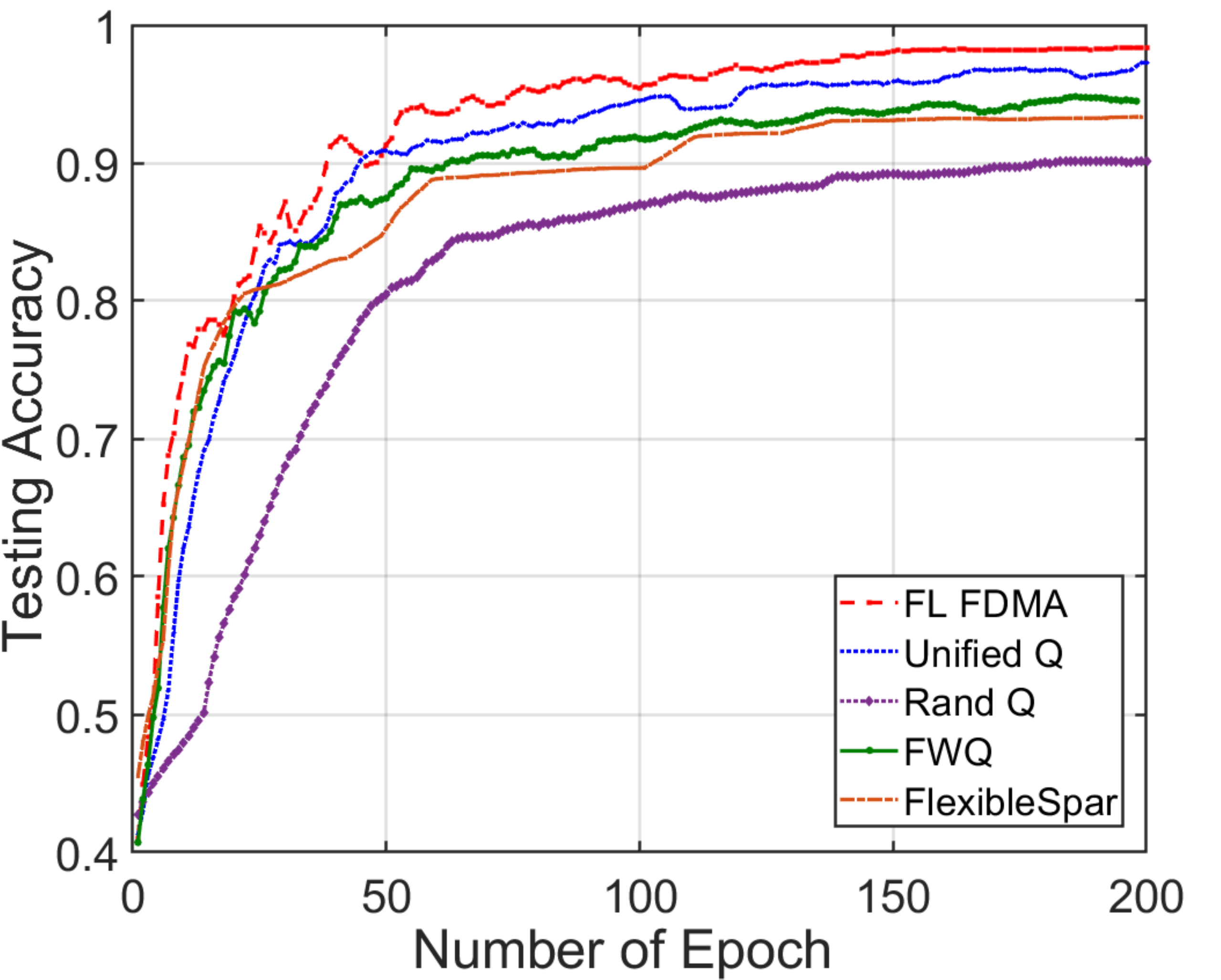}}
  \subfigure[Training error vs energy cost.\label{fig:EvACC} ]
  {\includegraphics[width=4.4cm]{ 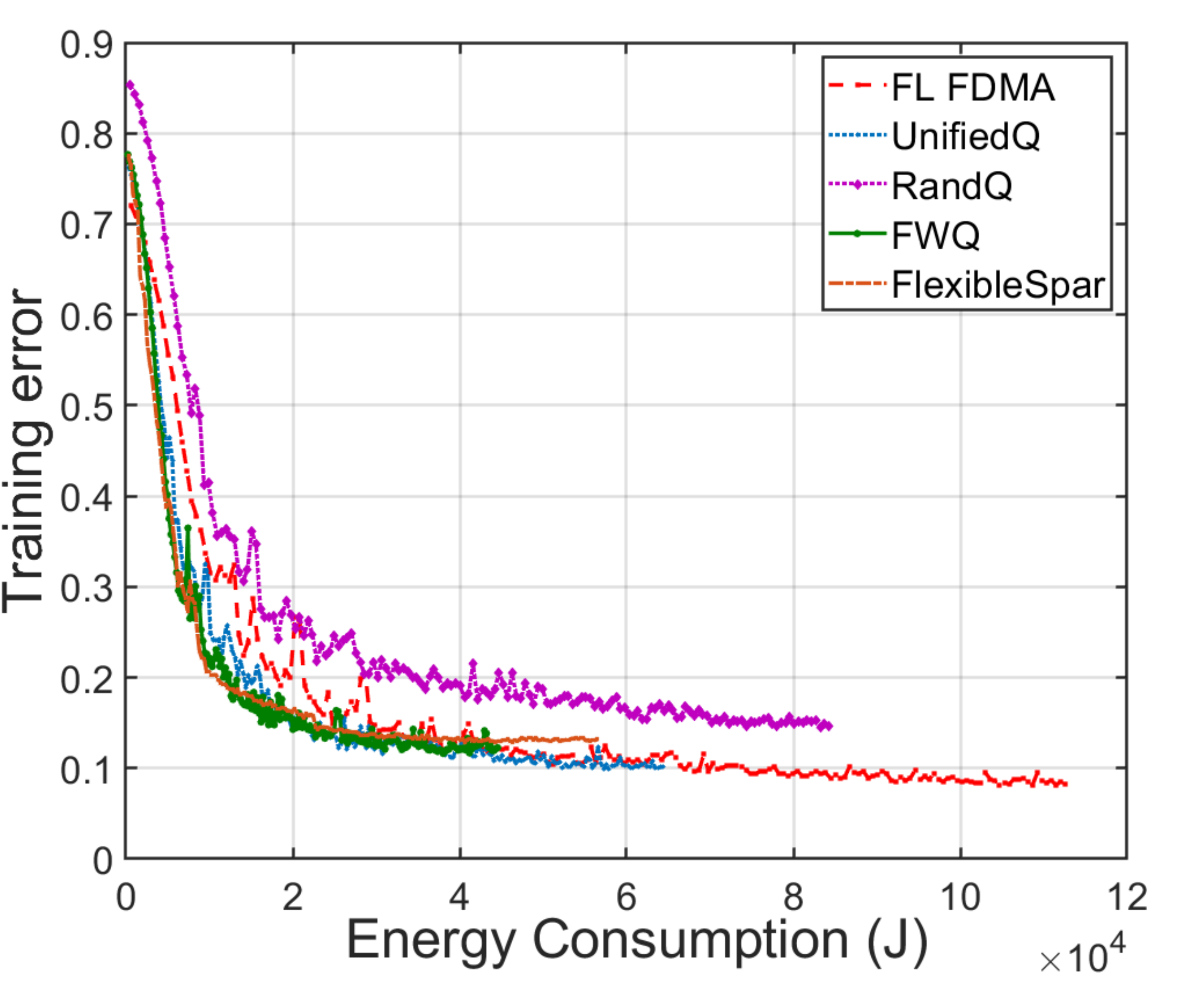}}
  \subfigure[Testing accuracy vs epochs.\label{fig:TvIcifar} ]
  {\includegraphics[width=4.4cm]{ 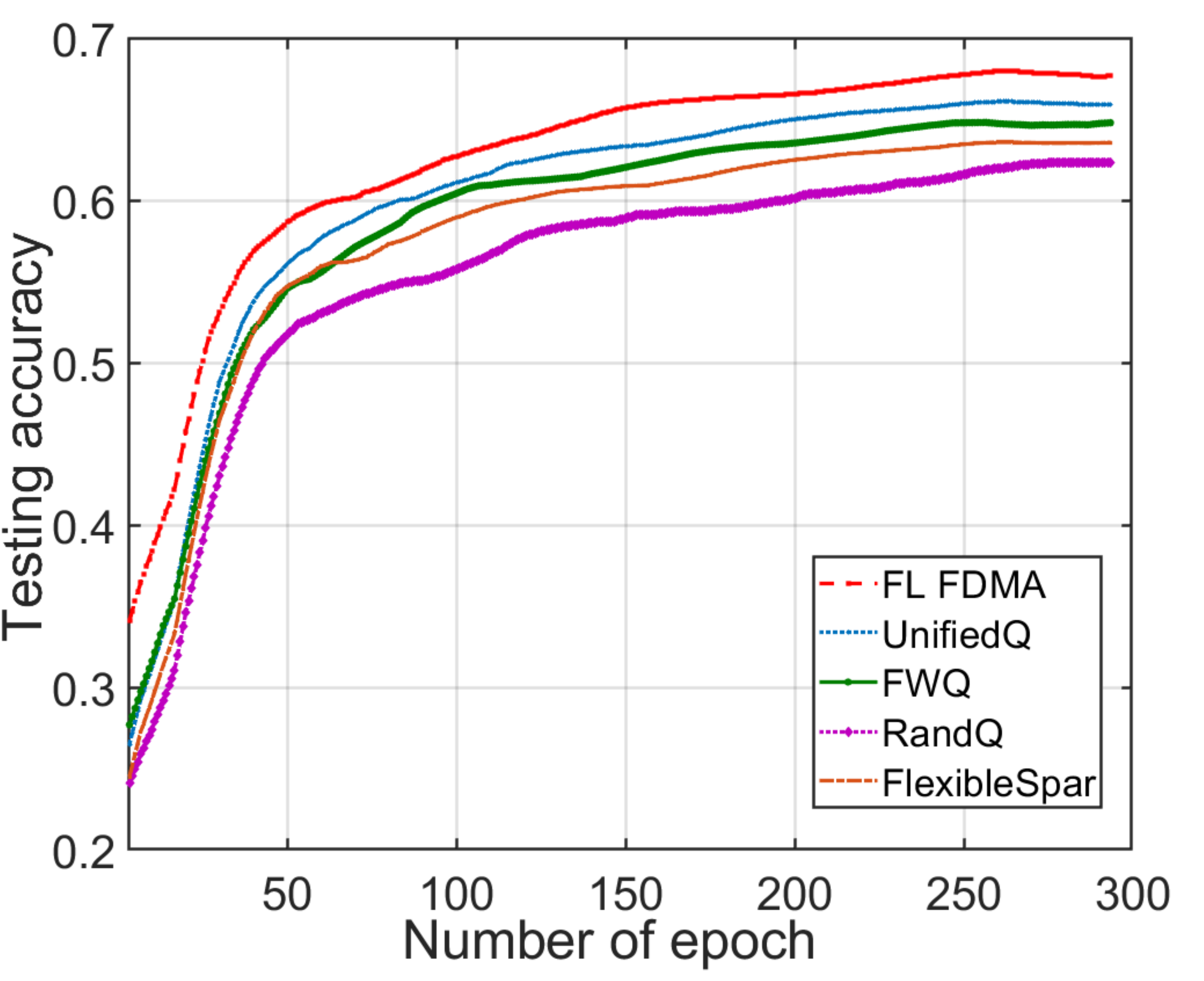}}
  \subfigure[Training error vs energy cost.\label{fig:EvACCcifar} ]
  {\includegraphics[width=4.4cm]{  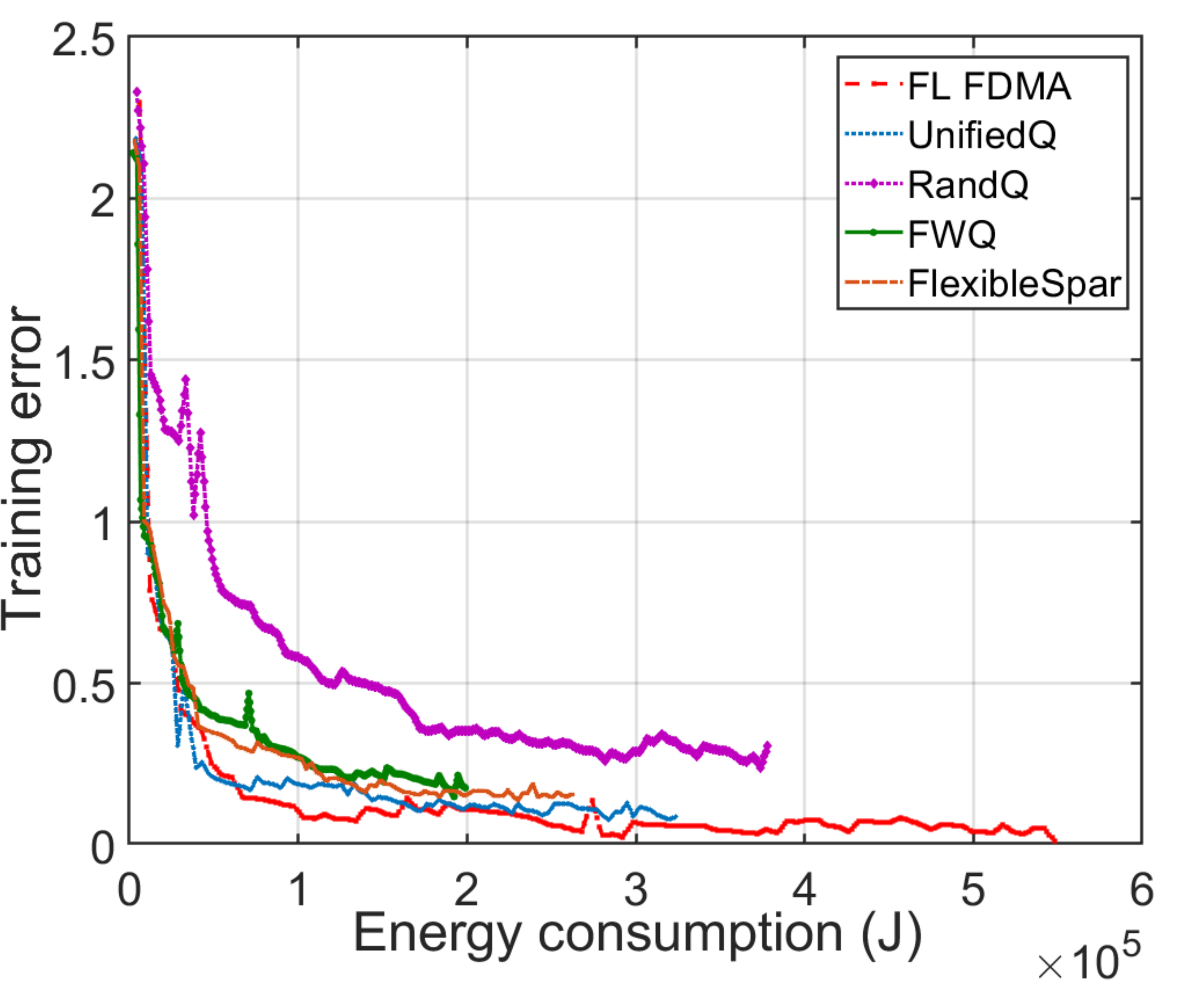}}
  \caption{Convergence Analysis. (a)-(b): ResNet-34 on CIFAR-10 with the estimated parameters $A_1=13.765, A_2=1.023, A_3=0.0435$ (c)-(d): MobileNet on CIFAR-100 with the estimated parameters $A_1=16.655, A_2=1.013, A_3 = 0.0795$} \label{Fig:sim}
\end{figure*}

\begin{figure*}[!htbp]
\centering
\begin{minipage}[b]{0.32\textwidth}
\centering
\includegraphics[width=5.4cm]{ 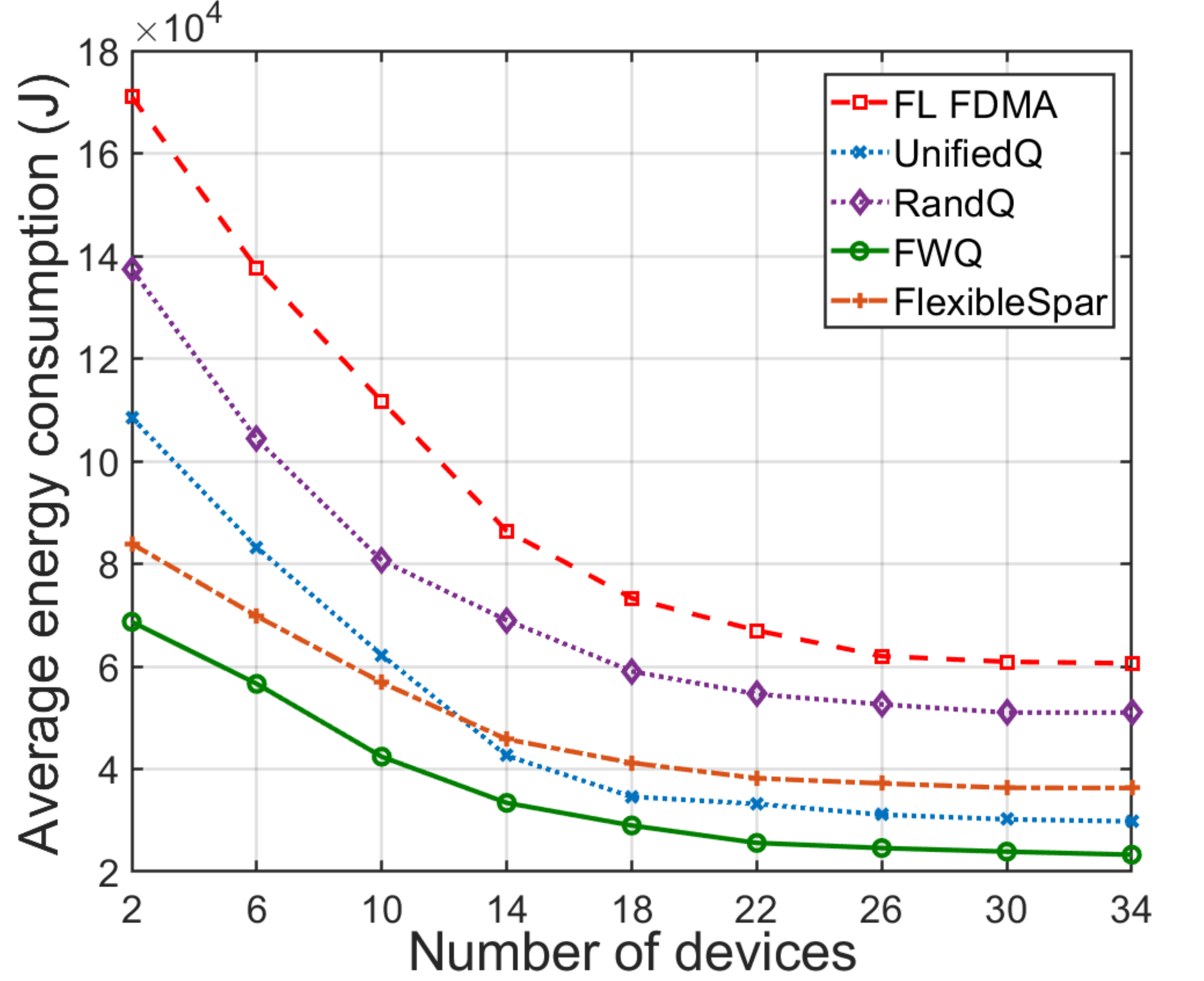} 
\caption{Energy vs the numbers of devices.}\label{fig:EvU}
\end{minipage}
\begin{minipage}[b]{0.32\textwidth}
\centering
\includegraphics[width=5.4cm]{ 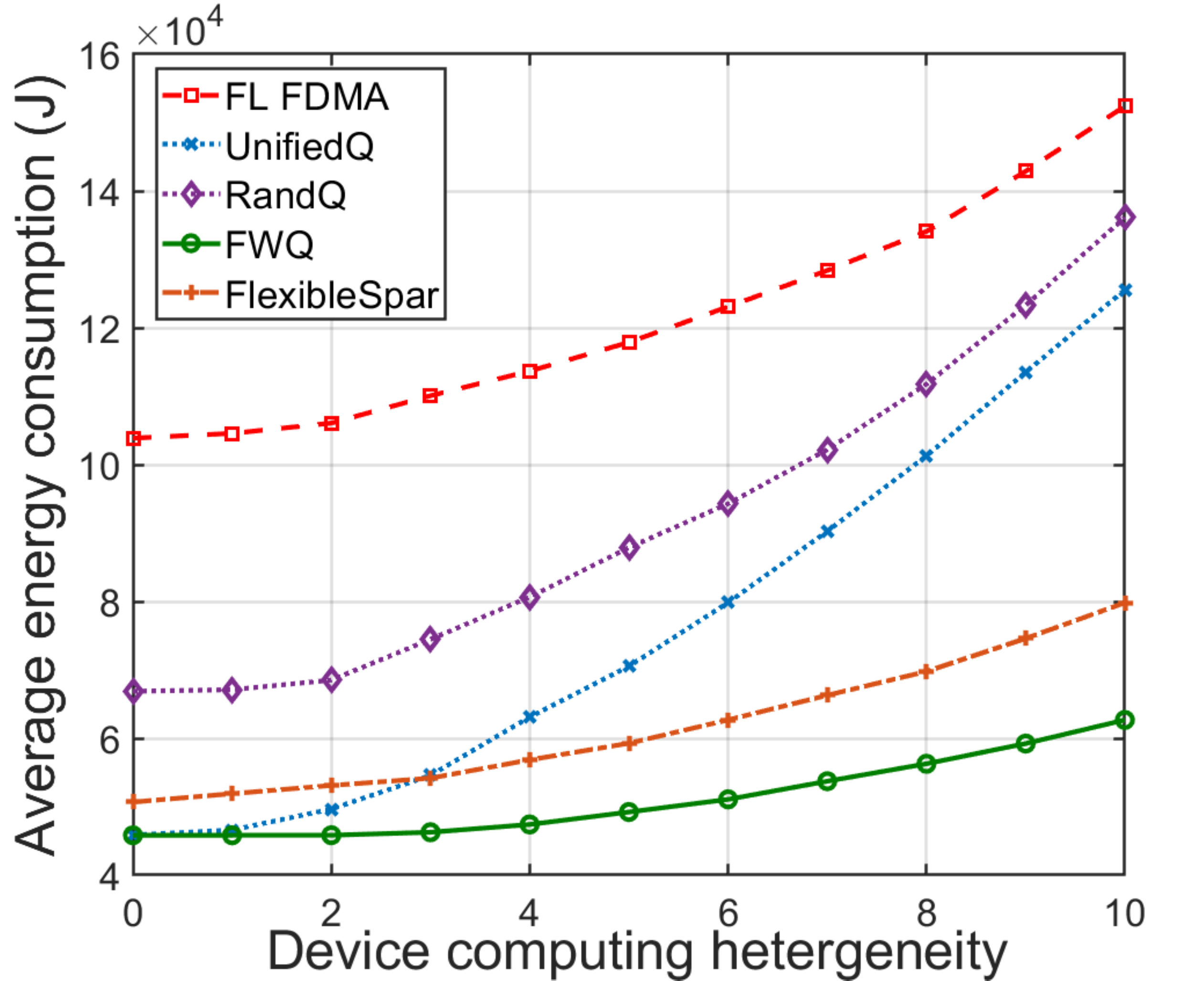}
\caption{Energy vs device heterogeneity.}\label{fig:EvH}
\end{minipage}
\begin{minipage}[b]{0.32\textwidth}
\centering
\includegraphics[width=5.4cm]{ 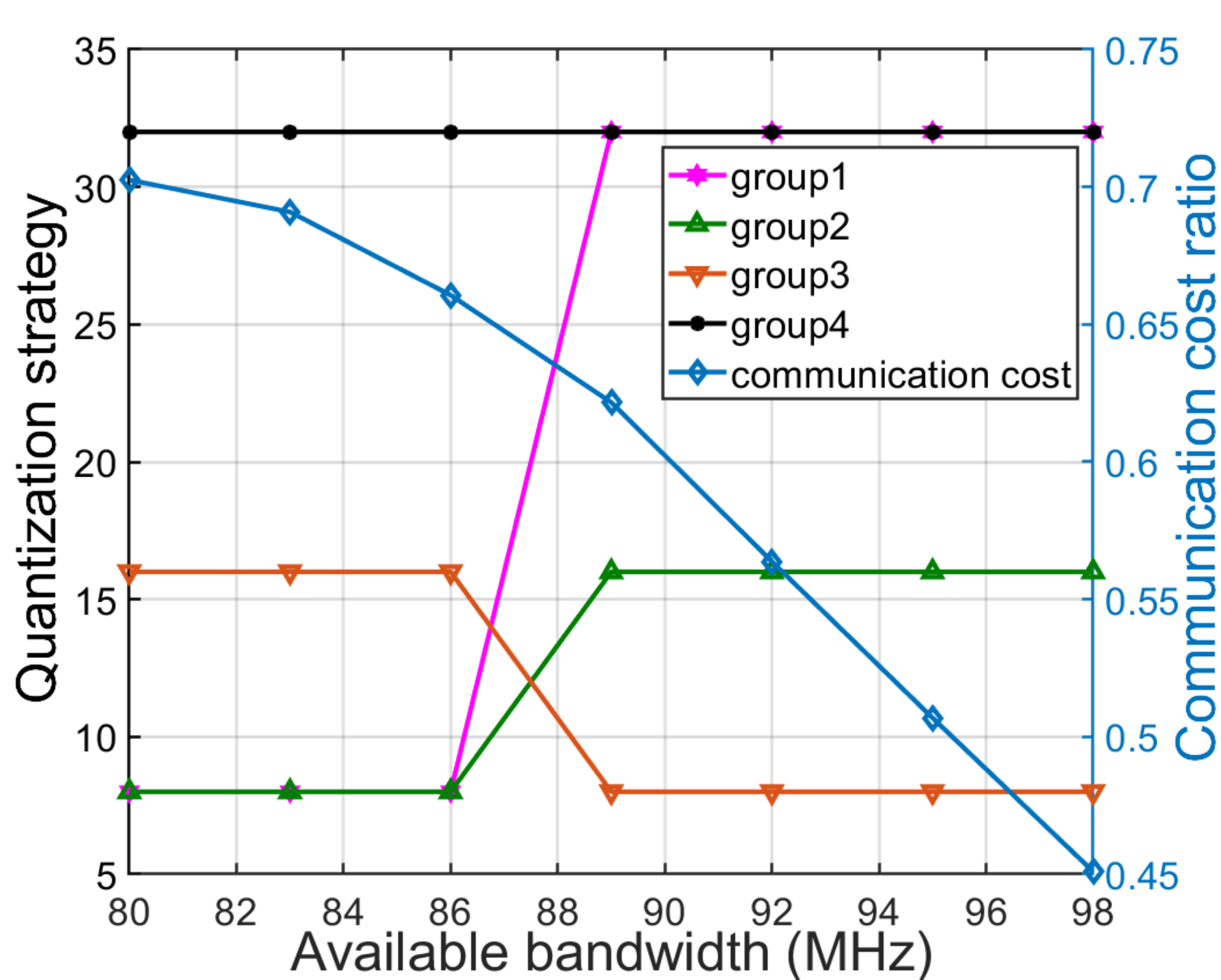}
\caption{Quantization vs bandwidth.}\label{fig:QvC} 
\end{minipage}
\end{figure*}

\section{Performance Evaluation}\label{Sec:Result}


\subsection{Data and settings}
1) \textit{Learning Model and Dataset:} To test the model performance, we choose two commonly-used deep learning models: ResNet-34 and Mobilenet. The well-known datasets, CIFAR-10 and CIFAR-100, are used to train FL models for image classification tasks. The CIFAR-10 dataset consists of 60000 32x32 color images in 10 classes with 5000 training images per class. The CIFAR-100 dataset has 100 classes and each class has 500 32x32  training images and 100 testing images. For the CIFAR-10 dataset, each device contains a  total number of $20000/N$ training samples with only 4 classes. For the CIFAR-100, each device contains a  total number of $20000/N$ training samples with only 40 classes. 

2) \textit{Communication and Computing Models: }For the communication model, we assume the noise power is $N_0 = -174 $ dBm. The transmitting power of each device is uniformly selected from $\{19, 20, 21, 22, 23\}$ dBm. Unless specified otherwise, we set the bandwidth $B_{\max}=100$MHz and the channel gains $h_i$ are modeled as i.i.d. Rayleigh fading with average path loss set to $10^{-3}$. Furthermore, we assume that model parameter is quantized into 16 bits before transmission. For the GPU computing model, the scaling factors of quantization are measured by Nvidia profiling tools on Jetson Xavier NX. We use ResNet-34 model with CIFAR-10 multiple times and obtain the simulated function $c_1(q) = 7.12\times10^{-3}q+0.274$ and $c_2(q) = 4.24\times10^{-4}q+1.035$. The GPU core frequency $f_i^{core}, \forall i$ is uniformly selected from $\{1050, 1100, 1150, 1200\}$MHz and memory frequency $f_i^{mem}, \forall i$ is uniformly selected from $\{1450, 1500, 1550, 1600\}$MHz.

3) \textit{Peer Schemes for Comparison:} We compare our proposed FWQ scheme with the following two different energy efficient FL schemes: 
\begin{itemize}
\item \textit{FL FDMA}~\cite{yang2020energy}: All mobile devices train their local models with full precision, i.e., without quantization. Their scheme optimizes the computing and communication resources (i.e., CPU frequency and wireless bandwidth) to minimize the energy consumption in FL training. For a fair comparison, we change the CPU model in \textit{FL FDMA} to GPU model and set $q=32$.
\item \textit{FlexibleSpar}~\cite{li2021talk}: All mobile devices train their local models with the full precision and sparsity of their model updates before transmitting to the FL server. Their scheme optimizes the frequency of model updates and gradient sparsity ratio to minimize the energy consumption in FL training. Here, we set $q=32$ in the GPU model.
\end{itemize}
Beside, we also consider two different quantization levels for our evaluation:
\begin{itemize}
\item \textit{Unified Q}: All the devices are set to use the same quantization strategy regardless of resource budgets for different mobile devices. 
\item \textit{Rand Q}: All mobile devices choose a quantization level randomly without considering the learning performance.
\end{itemize}
The resource allocation strategies for \textit{Unified Q} and \textit{Rand Q} are optimized by solving a simplified version of the problem (\ref{Eq:Reobj}).


\subsection{Convergence analysis}
First, we conduct convergence analysis. We implement the above learning models and choose a unified quantization strategy $q_1=\cdots=q_N=16$ in the “Unified Q” scheme. Fig.~\ref{fig:TvI} and Fig.~\ref{fig:TvIcifar} show the model performance in terms of testing accuracy. Here the testing accuracy of the model trained by “FL FDMA” is regarded as a baseline because “FL FDMA” trains the local models in full precision and transmits full gradients to the server. From Fig.~\ref{fig:TvI} and Fig.~\ref{fig:TvIcifar}, the weight quantization used in “FWQ” and “Unified Q” exhibits slight accuracy degradation. That is consistent with our convergence analysis that the discretization error induced by the quantization is unavoidable. This error is accumulated by all the participating mobile devices, which indicates some mobile devices take aggressive quantization levels (e.g., 8 bit) due to their resource limitation. As for our proposed FWQ scheme, since it considers this error in the quantization selection, the degradation is well controlled and relatively small. The corresponding energy consumption in the FL training is presented in Fig.~\ref{fig:EvACC} and Fig.~\ref{fig:EvACCcifar}. It demonstrates the effectiveness of our proposed scheme. The FWQ scheme is superior to the other three schemes in terms of the trade-off between the overall energy efficiency for FL training and training accuracy. It consumes x2 - x100 less energy than the other three schemes in the FL training process with only a 1.1\% accuracy loss.  

\subsection{Impact of number of users}
We now evaluate how the number of users affects the total energy consumption for FL training. Fig.~\ref{fig:EvU} shows that the average energy consumption decreases as the number of mobile devices participating in FL increases. The average energy consumption per device does not experience too much change even after more devices participate in FL training under all the schemes. The reason is that bringing more devices to train the FL model helps speed up the model convergence and thus reduce energy consumption, which is consistent with the sub-linear speedup in Theorem~\ref{tm:nomcvx_convergence}. However, as $N$ continues increasing, the marginal reduction of the total number of training iterations becomes smaller and smaller. Besides, our proposed FWQ scheme outperforms “FL FDMA” and “FlexibleSpar”. For example, the proposed FWQ scheme saves the energy of “FL FDMA” by 56\% and of “FlexibleSpar” by 35\%, respectively. The reason is that the proposed FWQ leverages weight quantization to reduce the workload for on-device training and optimize the weight quantization levels for heterogeneous devices, while the computing workload is not optimized and fixed for all the devices in both “FL FDMA” and “FlexibleSpar” scheme. Moreover, the proposed FWQ scheme reduces the energy in the “Unified Q” strategy by 20\% and the “Rand Q” strategy by 38.7\%, respectively, when the number of users is equal to 10. These results demonstrate the effectiveness of our proposed weight quantization scheme.



\subsection{Impact of computing capacities}
We evaluate the impact of device heterogeneity concerning computing capability. Here, we keep the number of mobile devices as ten and divide them into four groups. Fixing the minimum capacity as 1800MB, we set different capacities into 4 groups: $C$MB, $(C + 50L)$MB, $(C +150L)$MB, and $(C +200L)$MB, respectively. The values of $L$ range from 0 to 10. A larger value of $L$ means mobile devices have more diverse computing conditions, implying that the optimized quantization strategy has more diverse values. From Fig.~\ref{fig:EvH}, we observe that the total energy consumption grows as the value of $L$ increases. It indicates that device heterogeneity does impact the energy efficiency in FL training. Since the proposed FWQ scheme jointly optimizes the quantization levels and bandwidth allocation for heterogeneous devices, the FWQ scheme is superior to all other schemes in terms of high levels of computing heterogeneity across participating devices.


\subsection{Impact of communication capacities}
Figure~\ref{fig:QvC} shows the impacts of the wireless conditions on the optimal quantization selection. We vary the total available bandwidth from 80 MHz to 98 MHz and divide the mobile devices into 4 different groups, denoted as $\{g_1, g_2, g_3, g_4\}$, where the channel gain $h(g_1) \leq h(g_2) \leq h(g_d3) \leq h(g_4)$. From Fig.~\ref{fig:QvC}, we observe that, as the overall bandwidth becomes small, the ratio of the communication energy consumption to the overall energy consumption grows, which means wireless communications have a larger impact on the total energy consumption than local computing. As a result, the mobile devices in group 1, with small channel gain, become the stragglers in FL training and could slow down the gradient update time for one iteration.
To avoid the update delay for the next iteration and reduce the overall energy consumption, they have to take aggressive actions to compress their local models into the smallest number of bits. However this results in large discretization noise and degrades the performance, as stated in Theorem~\ref{tm:nomcvx_convergence}. To compensate for that, those who have better channel gain need to ``work" more by using a higher precision model to perform local training. Similarly, when the available bandwidth increases, the computing contributes more to the overall energy consumption. Those mobile devices with smaller local computing capacities choose to compress their models more to save computing energy.

\section{Conclusion}\label{Sec:con}
In this paper, we have studied the energy efficiency of FL training via joint design of wireless transmission and weight quantization. We have jointly exploited the flexible weight quantization selection and the bandwidth allocation to develop an energy efficient FL training algorithm over heterogeneous mobile devices, constrained by the training delay and learning performance. The weight quantization approach has been leveraged to deal with the mismatch between high model computing complexity and limited computing capacities of mobile devices. The convergence rate of FL with local quantization has been analyzed. Guided by the derived theoretical convergence bound, we have formulated the energy efficient FL training problem as a mixed-integer nonlinear programming. Since the optimization variables of the problem are strongly coupled, we have proposed an efficient iterative algorithm, where the closed-form solution of the bandwidth allocation and weight quantization levels are derived in each iteration. By comparing with different quantization levels through extensive simulations, we have demonstrated the effectiveness of our proposed scheme in handling device heterogeneity and reducing overall energy consumption in FL over heterogeneous mobile devices.

\bibliographystyle{IEEEtran}      
\bibliography{ref_rui}

\clearpage
\newpage

\appendices
\section{Proof of Theorem~\ref{tm:nomcvx_convergence}}\label{PF:T1}
\subsection{Additional Notation}
For the simplicity of notations, we denote the error of weight quantization ${\boldsymbol{e}}_i^{r} \triangleq Q_i\left({\boldsymbol{w}}_i^{r+1}\right) - {\boldsymbol{w}}_i^{r+1}$, and the local “gradients” with weight quantization as $\widetilde{\boldsymbol{g}}_i^{r} \triangleq \triangledown \widetilde{f}_i ({\boldsymbol{w}}_i^{r} ) - {\boldsymbol{e}}_i^{r}/\eta$. Since the quantization scheme $Q_i(\cdot)$ we used is an unbiased scheme, $\mathbb{E}_Q\left[{\boldsymbol{e}}_i^{r}\right] = 0$. 


Inspired by the iterate analysis framework in~\cite{li2020federated}, we define the following virtual sequences:
\begin{flalign}
{\boldsymbol{u}}_i^{r+1} &= {\boldsymbol{w}}_i^{r} - \eta \widetilde{\boldsymbol{g}}_i^{r}, \\
{\boldsymbol{w}}_i^{r+1} &= \kappa^{r} \sum_{i=1}^N \pi_i  {\boldsymbol{u}}^{r+1}_i + (1- \kappa^{r}){\boldsymbol{u}}_i^{r+1}\label{def:w}, 
\end{flalign}
where $\kappa^{r} = 1$ if $(r+1) \mod H = 0$ and $\kappa^{r} = 0$ otherwise.
The following short-hand will be found useful in the convergence analysis of the proposed FWQ framework 
\begin{flalign}
    &\bar{\boldsymbol{u}}^{r}  = \sum_{i=1}^N \pi_i  {\boldsymbol{u}}_i^{r}, \quad \bar{\boldsymbol{w}}^{r} = \sum_{i=1}^N \pi_i  {\boldsymbol{w}}_i^{r},  \quad
    \widetilde{\boldsymbol{g}}^{r} =\sum_{i=1}^N \pi_i  \widetilde{\boldsymbol{g}}_i^{r}.\label{notation}
\end{flalign}
Thus, $\bar{\boldsymbol{u}}^{r+1} = \bar{\boldsymbol{w}}^{r} - \eta \widetilde{\boldsymbol{g}}^{r}$. Note that we can only obtain $\bar{\boldsymbol{w}}^{r+1}$ when $(r+1) \mod H = 0$. 

\subsection{Key Lemmas}
Now, we give four important lemmas to offer our proof.

\begin{lemma}[Weight quantization error~\cite{li2017training}\label{l:e_r}]
If Assumption~\ref{as:divergence} holds, the stochastic rounding error on each iteration can be bounded, in expectation
\begin{flalign}
     %
     &\mathbb{E}_Q\left[\norm{{\boldsymbol{e}}_i^{r} }_2^2\right] \leq \eta \sqrt{d} G \delta_i.
\end{flalign}
\end{lemma}
 
\begin{lemma}[Bounding the divergence\label{eq:Bound}]
Suppose $1-3\eta^2 L^2 H^2 > 0$, we have,
\begin{flalign}
&\sum_{r=0}^{R-1} \sum_{i=1}^N \pi_i^2  \mathbb{E}\left[\mathbb{E}_Q\left[\norm{ \bar{\boldsymbol{w}}^{r}-{\boldsymbol{w}}_i^{r} }_2^2\right]\right] \nonumber\\
& \leq \frac{ \eta^2 RH \tau}{M(1-3\eta^2 L^2 H^2)}  + \frac{\eta \sqrt{d} RH G \sum_{i=1}^N \pi_i^2 \delta_i }{1-3\eta^2 L^2 H^2} &\nonumber\\
 &\quad + \frac{3\eta^2 H^2 \pi}{1-3\eta^2 L^2 H^2 }\sum_{r=0}^{R-1} \norm{ \triangledown {F} ( \bar{\boldsymbol{w}}^{r}) }_2^2,&
\end{flalign}
where $\tau = \sum_{i=1}^N  \pi_i^2\tau_i^2$ and $\pi = \sum_{i=1}^N \pi_i^2$. 
 
\begin{proof}
Recalling that at the synchronization step $r' \in \mathcal{U}_H$, $\boldsymbol{w}_i^{r'}=\bar{\boldsymbol{w}}^{r'}$ for all $i \in \mathcal{N}$. Therefore, for any $r \geq 0$, such that for $r' \leq r \leq r'+H$, we have
\begin{flalign}
    &A1_r:=\sum_{i=1}^N \pi_i^2  \mathbb{E}\left[\mathbb{E}_Q\left[\norm{ \bar{\boldsymbol{w}}^{r} - {\boldsymbol{w}}_i^{r} }_2^2\right]\right] & \nonumber\\
    &= \sum_{i=1}^N \pi_i^2 \mathbb{E} \left[\mathbb{E}_Q \left[ \lVert{ ({\boldsymbol{w}}_i^{r}-\bar{\boldsymbol{w}}^{r'}) - (\bar{\boldsymbol{w}}^{r} -\bar{\boldsymbol{w}}^{r'})   }\rVert_2^2\right]\right] \nonumber\\
    &\overset{(a)}{\leq} \sum_{i=1}^N \pi_i^2 \mathbb{E} \left[\mathbb{E}_Q \left[\lVert{ {\boldsymbol{w}}_i^{r}- {\boldsymbol{w}}_i^{r'} }\rVert_2^2  \right]  \right] &\nonumber\\
    &= \sum_{i=1}^N \pi_i^2  \mathbb{E}  [\lVert{\sum_{j=r'}^{r} -\eta \triangledown \widetilde{f}_i ({\boldsymbol{w}}_i^{j})+ \boldsymbol{e}_i^r }\rVert_2^2 ]& \nonumber\\
    &= \sum_{i=1}^N \pi_i^2  \mathbb{E} [\lVert{\sum_{j=r'}^{r}\eta \triangledown \widetilde{f}_i ({\boldsymbol{w}}_i^{j} )}\rVert_2^2 +  \sum_{j=r'}^{j} \mathbb{E}_Q  [\norm{\boldsymbol{e}_i^r }_2^2]] & \nonumber\\
    &\overset{(b)}{\leq} \underbrace{\eta^2 \sum_{i=1}^N  \pi_i^2  \mathbb{E} [ \lVert{\sum_{i=r'}^{r'_H} \triangledown \widetilde{f}_i ({\boldsymbol{w}}_i^{j} ) } \rVert_2^2 ]}_{A_2} +  \eta \sqrt{d} H G \sum_{i=1}^N \pi_i^2 \delta_i, \label{eq:A1}  
\end{flalign}
where $r'_H = r'+H-1$, $(a)$ results from $\sum_{i=1}^{n} \pi_i ({\boldsymbol{w}}_i^{r}- {\boldsymbol{w}}_i^{r'}) = \bar{\boldsymbol{w}}^{r} -\bar{\boldsymbol{w}}^{r'}$, ${\boldsymbol{w}}_i^{r'} = \bar{\boldsymbol{w}}^{r'}$, and $\mathbb{E} \lVert{\boldsymbol{x} - \mathbb{E}[\boldsymbol{x}]} \rVert_2^2 \leq \mathbb{E}\lVert{\boldsymbol{x} }\rVert_2^2$. 
$(b)$ follows Lemma 1. We generalize the result from \cite{jiang2018linear} to upper-bound the second term $A_2$ in RHS of (\ref{eq:A1}):
\begin{flalign}
     A_2&:= \eta^2 \sum_{i=1}^N  \pi_i^2 \mathbb{E} [\lVert{\sum_{j=r'}^{r'_H} \triangledown \widetilde{f}_i ({\boldsymbol{w}}_i^{j})- \triangledown{F}_i ({\boldsymbol{w}}_i^{j})+ \triangledown{F}_i ({\boldsymbol{w}}_i^{j}) }\rVert_2^2 ] \nonumber\\
     &\leq 3\eta^2  H \sum_{i=1}^N \pi_i^2 \sum_{j=r'}^{r'_H} \left( L^2 \norm{ \bar{\boldsymbol{w}}^{j}- {\boldsymbol{w}}_i^{j}}_2^2  +  \norm{ \triangledown {F} ( \bar{\boldsymbol{w}}^{j}) }_2^2\right) \nonumber\\
     &\quad + \eta^2H \frac{\sum_{i=1}^N \pi_i^2\tau_i^2 }{M} .  \label{eq:A2}  
\end{flalign}
It follows that
\begin{flalign}
     \sum_{r=0}^{R-1} A1_r &\leq \eta^2 RH \frac{\tau }{M}  + 3\eta^2 H^2 \sum_{r=0}^{R-1} ( L^2{A1}_r + \pi\norm{ \triangledown {F} ( \bar{\boldsymbol{w}}^{r}) }_2^2 )  \nonumber\\
     &\quad  + \eta \sqrt{d} H G \sum_{i=1}^N \pi_i^2 \delta_i,
\end{flalign}
where $\tau = \sum_{i=1}^N  \pi_i^2 \tau_i^2$ and $\pi =  \sum_{i=1}^N \pi_i^2$. 

Suppose $1-3\eta^2 L^2 H^2 > 0$, we have
\begin{flalign}
\sum_{r=0}^{R-1} A1_r& \leq \frac{ \eta^2 RH \tau }{M(1-3\eta^2 L^2 H^2)} + \frac{\eta \sqrt{d} H G \sum_{i=1}^N \pi_i^2 \delta_i }{1-3\eta^2 L^2 H^2 }& \nonumber\\
&\quad + \frac{3\eta^2 H^2 \pi}{1-3\eta^2 L^2 H^2 }\sum_{r=0}^{R-1} \norm{ \triangledown {F} ( \bar{\boldsymbol{w}}^{r}) }_2^2, \label{eq:suma}&
\end{flalign}
and the proof is completed.
\end{proof}
\end{lemma}

\begin{lemma}\label{l:inner}
According to the proposed algorithm the expected inner product can be bounded with:
\begin{flalign}
&\mathbb{E}  \left[\mathbb{E}_Q \left[  {\langle{ \triangledown F(\bar{\boldsymbol{w}}^{r}),  \bar{\boldsymbol{u}}^{r+1} - \bar{\boldsymbol{w}}^{r} } \rangle} \right] \right]   \nonumber  \\
&\leq - \frac{\eta}{2} \mathbb{E}  \left[\norm{\triangledown F(\bar{\boldsymbol{w}}^{r})}_2^2\right] + \frac{\eta L^2}{2} \sum_{i=1}^N \pi_i^2 \mathbb{E}\left[\mathbb{E}_Q\left[\norm{ \bar{\boldsymbol{w}}^{r}-{\boldsymbol{w}}_i^{r} }_2^2\right]\right] \nonumber \\
&\quad  - \frac{\eta}{2} \mathbb{E}  \left[ \norm{\sum_{i=1}^N \pi_i \triangledown F_i({\boldsymbol{w}}_i^{r})}_2^2\right].
\end{flalign}

\begin{proof}
\begin{flalign}
    &\mathbb{E}  \left[  {\langle{ \triangledown F(\bar{\boldsymbol{w}}^{r}),  \bar{\boldsymbol{u}}^{r+1} - \bar{\boldsymbol{w}}^{r} } \rangle}  \right] & \nonumber\\
     &= \mathbb{E}  \left[ { \left \langle { \triangledown F(\bar{\boldsymbol{w}}^{r}),- \eta \widetilde{\boldsymbol{g}}^{r}} \right \rangle}  \right]& \nonumber\\
     &= - \eta{ \left \langle { \triangledown F(\bar{\boldsymbol{w}}^{r}), \mathbb{E}  \left[ \sum_{i=1}^N \pi_i  \triangledown {F}_i ({\boldsymbol{w}}_i^{r} )  \right] } \right \rangle}&  \nonumber\\
     &\overset{(a)}{=} -\frac{\eta}{2} \mathbb{E}  \left[\norm{\triangledown F(\bar{\boldsymbol{w}}^{r})}_2^2\right] - \frac{\eta}{2}   \mathbb{E}  \left[\norm{\sum_{i=1}^N \pi_i  \triangledown F_i({\boldsymbol{w}}_i^{r})}_2^2\right]&  \nonumber\\
     &\quad+ \frac{\eta}{2} \mathbb{E}  \left[\norm{\triangledown F(\bar{\boldsymbol{w}}^{r}) -  \sum_{i=1}^N \pi_i \triangledown F_i({\boldsymbol{w}}_i^{r})}_2^2\right]& \nonumber\\
     &\overset{(b)}{\leq} - \frac{\eta}{2} \mathbb{E}  \left[\norm{\triangledown F(\bar{\boldsymbol{w}}^{r})}_2^2\right] - \frac{\eta}{2} \mathbb{E}  \left[ \norm{\sum_{i=1}^N \pi_i \triangledown F_i({\boldsymbol{w}}_i^{r})}_2^2\right] &  \nonumber  \\
     &\quad+ \frac{\eta L^2}{2}   \sum_{i=1}^N \pi_i^2 \mathbb{E}\left[\mathbb{E}_Q\left[\norm{ \bar{\boldsymbol{w}}^{r}-{\boldsymbol{w}}_i^{r} }_2^2\right]\right] ,&
\end{flalign}
where $(a)$ is due to $2\langle a,b \rangle = ||a||^2+||b||^2-||a-b||^2$ and $(b)$ follows from $L$-smoothness assumption.
\end{proof}
\end{lemma}

\begin{lemma}\label{l:n2}
If Assumptions~\ref{as:Lsmooth} and~\ref{as:divergence} hold, then for sequences $\bar{\boldsymbol{w}}$ defined in (\ref{def:w})-(\ref{notation}), we have
\begin{flalign}
    &\frac{ L}{2} \mathbb{E} \left[\norm{\bar{\boldsymbol{w}}^{r+1}-\bar{\boldsymbol{w}}^{r}}_2^2 \right]\nonumber\\
    &\leq \frac{\eta^2 L}{2} \norm{\sum_{i=1}^N \pi_i \triangledown {F}_i ({\boldsymbol{w}}_i^{r} )}_2^2+  \frac{\eta^2 L\tau  }{2M} + \frac{\eta L}{2} \sqrt{d} G \sum_{i=1}^N \pi_i^2 \delta_i \nonumber\\ 
     & \quad + \frac{ L}{2 }\sum_{i=1}^N  \mathbb{E} \left[
     \norm{\kappa^{r} \pi_i\left({\boldsymbol{\Delta}}^{r+1}_i - \bar{\boldsymbol{\Delta}}^{r+1} \right)}_2^2
     \right],
\end{flalign}
where ${\boldsymbol{\Delta}}^{r+1}_i = {\boldsymbol{w}}_i^{r'} - {\boldsymbol{u}}_i^{r+1}$ and $\bar{\boldsymbol{\Delta}}^{r} = \sum_{i=1}^N \pi_i{\boldsymbol{\Delta}}^{r+1}_i$. 

\begin{proof}
According to the updating rules defined in (\ref{def:w}), we have,
\begin{flalign}
&\frac{ L}{2} \mathbb{E} \left[\norm{\bar{\boldsymbol{w}}^{r+1}-\bar{\boldsymbol{w}}^{r}}_2^2 \right] \nonumber\\
&=\frac{ L}{2} \mathbb{E} [
\lVert{\sum_{i=1}^N \pi_i {\boldsymbol{u}}_i^{r+1}-{\boldsymbol{w}}_i^{r} + \kappa^{r} \left({\boldsymbol{w}}_i^{r'} - \bar{\boldsymbol{\Delta}}^{r} - {\boldsymbol{u}}_i^{r+1} \right)}\rVert_2^2
]\nonumber
\end{flalign}
\begin{flalign}
&= \frac{ L}{2}\mathbb{E} \left[\eta^2 \norm{ \hat{\boldsymbol{g}}^{r}}_2^2  + \sum_{i=1}^N  \pi_i^2 
\norm{\kappa^{r}\left({\boldsymbol{\Delta}}^{r+1}_i - \bar{\boldsymbol{\Delta}}^{r+1} \right)}_2^2
\right],\label{eq:A3}
\end{flalign}
where ${\boldsymbol{\Delta}}^{r+1}_i = {\boldsymbol{w}}_i^{r'} - {\boldsymbol{u}}_i^{r+1}$ and $\bar{\boldsymbol{\Delta}}^{r+1} = \sum_{i=1}^N \pi_i{\boldsymbol{\Delta}}^{r+1}_i$. The first term in (\ref{eq:A3}) can be bounded by,
\begin{flalign}
 &\frac{\eta^2 L}{2}\mathbb{E} \left[\norm{\sum_{i=1}^N \pi_i \triangledown \widetilde{f}_i ({\boldsymbol{w}}_i^{r} ) -{\boldsymbol{e}}_i^{r}/\eta}_2^2 \right]  &\nonumber\\
 &=\frac{\eta^2 L}{2}\mathbb{E} \left[\norm{ \sum_{i=1}^N \pi_i (\triangledown \widetilde{f}_i ({\boldsymbol{w}}_i^{r} ) - \triangledown {F}_i ({\boldsymbol{w}}_i^{r} )+\triangledown {F}_i ({\boldsymbol{w}}_i^{r} )) }_2^2 \right]&\nonumber\\
 &\quad + \frac{L}{2 } \sum_{i=1}^N \pi_i^2\mathbb{E} \left[\norm{{\boldsymbol{e}}_i^{r}}_2^2 \right]&\nonumber\\
  &\leq \frac{\eta  L}{2} ( \frac{\eta \tau}{ M} + \eta ||{\sum_{i=1}^N \pi_i \triangledown {F}_i ({\boldsymbol{w}}_i^{r} )}||_2^2 +  \sqrt{d}G \sum_{i=1}^N \pi_i^2 \delta_i ). &
  %
\end{flalign}
Since $\kappa^{r} = 0$ when $(r +1) \mod H \neq 0$, and we have the second term in (\ref{eq:A3}) equals to zero. When $(r +1) \mod H = 0$, 
\begin{flalign}
    &\frac{ L}{2}\sum_{i=1}^N \pi_i^2\mathbb{E}  \left[
     \norm{ \left({\boldsymbol{\Delta}}^{r+1}_i - \bar{\boldsymbol{\Delta}}^{r+1} \right)}_2^2\right]&\nonumber\\
     &\leq \frac{ L}{2}\sum_{i=1}^N \pi_i^2\mathbb{E} \left[
     \norm{ {\boldsymbol{w}}_i^{r'} - {\boldsymbol{u}}_i^{r+1}}_2^2\right]\nonumber &\\
     &= \frac{ L}{2}\sum_{i=1}^N  \pi_i^2 \mathbb{E}  [\lVert{ \sum_{j=r'}^{r} \eta\triangledown \widetilde{f}_i ({\boldsymbol{w}}_i^{r} ) - {\boldsymbol{e}}_i^{r}}\rVert_2^2 ]& \nonumber\\
     &\leq \frac{LA_2}{2} + \frac{\eta L H}{2} \sqrt{d} G \sum_{i=1}^N \pi_i^2 \delta_i  &\nonumber\\
     &\overset{(a)}{\leq}  \frac{3\eta^2LH\pi}{2}  \sum_{j=r'}^{r'_H} \norm{ \triangledown {F} ( \bar{\boldsymbol{w}}^{j}) }_2^2 + \frac{\eta L H}{2} \sqrt{d} G \sum_{i=1}^N \pi_i^2 \delta_i &\nonumber\\
     &\quad  + \frac{\eta^2LH \tau }{2M}  + \frac{3\eta^2 L^3 H}{2} \sum_{j=r'}^{r'_H}A1_j ,&
\end{flalign}
where $(a)$ follows the results in (\ref{eq:A2}).
\end{proof}
\end{lemma}
 
\subsection{Main Results}
Under the Lipschitzan gradient assumption on $F$, we have,
\begin{flalign}
     &\mathbb{E} \left[ F(\bar{\boldsymbol{w}}^{r+1})\right] - \mathbb{E} \left[F(\bar{\boldsymbol{w}}^{r}) \right]  \nonumber\\
     &\leq \mathbb{E} \left[  {\langle  {\triangledown F(\bar{\boldsymbol{w}}^{r}), \bar{\boldsymbol{w}}^{r+1}-\bar{\boldsymbol{w}}^{r}} \rangle} \right] + \frac{L}{2} \mathbb{E} \left[ \norm{\bar{\boldsymbol{w}}^{r+1}-\bar{\boldsymbol{w}}^{r}}_2^2 \right] \nonumber\\
     &= \mathbb{E}  \left[{\langle  {\triangledown F(\bar{\boldsymbol{w}}^{r}), \bar{\boldsymbol{w}}^{r+1}-\bar{\boldsymbol{u}}^{r+1} + \bar{\boldsymbol{u}}^{r+1}-\bar{\boldsymbol{w}}^{r}} \rangle}\right] \nonumber\\
     &\quad + \frac{ L}{2} \mathbb{E} \left[\norm{\bar{\boldsymbol{w}}^{r+1}-\bar{\boldsymbol{w}}^{r}}_2^2 \right] \nonumber\\ 
     &\overset{(a)}{=} \mathbb{E}\left[ {\langle{ \triangledown F(\bar{\boldsymbol{w}}^{r}),  \bar{\boldsymbol{u}}^{r+1} - \bar{\boldsymbol{w}}^{r} } \rangle}  +  \frac{L}{2} \norm{\bar{\boldsymbol{w}}^{r+1}-\bar{\boldsymbol{w}}^{r}}_2^2\right],
     \label{eq:n1} 
\end{flalign}
where $(a)$ is $\mathbb{E} [\mathbb{E}_Q [\bar{{\boldsymbol{w}}}^{r+1} ] ] = \mathbb{E}[ \bar{{\boldsymbol{u}}}^{r+1}]$. We use Lemma~\ref{l:e_r}-\ref{l:inner},\ref{l:n2} to upper bound two terms in the RHS of (\ref{eq:n1}) and obtain,
\begin{flalign}
    &\mathbb{E} \left[ F(\bar{\boldsymbol{w}}^{r+1})\right] - \mathbb{E} \left[F(\bar{\boldsymbol{w}}^{r}) \right]  \nonumber\\
    &\leq - \frac{\eta}{2} \mathbb{E}  \left[\norm{\triangledown F(\bar{\boldsymbol{w}}^{r})}_2^2\right]   +\left(-\frac{\eta}{2}+\frac{\eta^2L}{2}\right)\norm{\sum_{i=1}^N \pi_i \triangledown {F}_i ({\boldsymbol{w}}_i^{r} )}_2^2 \nonumber \\
    & \quad+ \frac{\eta L^2}{2} A1_r +  \frac{\eta^2 L \tau }{2M} + \frac{\eta L}{2} \sqrt{d} G \sum_{i=1}^N \pi_i^2 \delta_i\nonumber\\
    & \quad + \frac{ L}{2}\sum_{i=1}^N\mathbb{E} \left[
     \norm{\kappa^{r}  \pi_i  \left({\boldsymbol{\Delta}}^{r+1}_i - \bar{\boldsymbol{\Delta}}^{r+1} \right)}_2^2
     \right] \nonumber\\
    &\overset{(a)}{\leq} - \frac{\eta}{2} \mathbb{E}  \left[\norm{\triangledown F(\bar{\boldsymbol{w}}^{r})}_2^2\right]  + \frac{\eta L^2}{2} A1_r   + \frac{\eta L}{2} \sqrt{d} G \sum_{i=1}^N \pi_i^2 \delta_i \nonumber\\ 
    & \quad  +  \frac{\eta^2 L \tau }{2M} + \frac{ L}{2}\sum_{i=1}^N \mathbb{E} \left[
     \norm{\kappa^{r} \left({\boldsymbol{\Delta}}^{r+1}_i - \bar{\boldsymbol{\Delta}}^{r+1} \right)}_2^2
     \right].
\end{flalign}
Here, (a) holds if the learning rate $\eta L \leq 1$. Summing up for all $R$ communication rounds, we have: 
\begin{flalign}
     &\mathbb{E} \left[F(\bar{\boldsymbol{w}}^{R}) \right] - \mathbb{E} \left[F(\bar{\boldsymbol{w}}^{0})\right] & \nonumber\\
    %
    %
    &\leq \frac{\eta}{2}(3\eta L H-1) \sum_{r=0}^{R-1} \mathbb{E}  \norm{\triangledown F(\bar{\boldsymbol{w}}^{r})}_2^2 + \eta LR\sqrt{d} G \sum_{i=1}^N \pi_i^2 \delta_i & \nonumber\\ 
    & \quad  +  \frac{\eta^2 L R\tau }{ M}+ \frac{\eta L^2}{2} (1 + 3\eta L H) \sum_{r=0}^{R-1}A1_r.  \label{eq:r}  &
\end{flalign}
 
Plugging (\ref{eq:suma}) into (\ref{eq:r}), we can obtain
\begin{flalign}
    &\mathbb{E} \left[F(\bar{\boldsymbol{w}}^{R}) \right] - \mathbb{E} \left[F(\bar{\boldsymbol{w}}^{0})\right] & \nonumber\\
    &\leq -\frac{\eta}{2}(1 - 3\eta L H - \frac{3\eta^2L HC_1}{2}) \sum_{r=0}^{R-1} \mathbb{E}  \left[\norm{\triangledown F(\bar{\boldsymbol{w}}^{r})}_2^2\right] & \nonumber\\ 
    & \quad  +  R\frac{\eta^2 L \tau }{M } (1 + \frac{C_1}{2})  +  \frac{\eta L\sqrt{d}R G}{2} (\frac{1}{2} + C_1) \sum_{i=1}^N \pi_i^2 \delta_i,&
\end{flalign}
where $C_1 = \frac{\eta L H+3\eta^2 L^2 H^2\pi}{ 1-3\eta^2 L^2 H^2 }$
and rearranging the terms gives
\begin{flalign}
    & \frac{\eta}{2}C'_1  \sum_{r=0}^{R-1} \norm{ \triangledown {F} ( \bar{\boldsymbol{w}}^{r}) }_2^2 \nonumber\\ 
    &\leq \mathbb{E} \left[F(\bar{\boldsymbol{w}}^{0}) -F(\bar{\boldsymbol{w}}^{R})\right] + \frac{\eta L\sqrt{d}R G}{2}(1 + 2C_1) \sum_{i=1}^N \pi_i^2 \delta_i \nonumber\\ 
    &\quad    +  R\frac{\eta^2 L  }{2M}(2+ C_1) \tau, 
\end{flalign}
where $C'_1 =1 - 3\eta L H - \frac{3\eta^2L HC_1}{2}$.

If we set $\eta = \sqrt{M/R}$ and $    3\eta L H  \geq  \frac{3\eta^2L H(\eta L H+3\eta^2 L^2 H^2\pi)}{2(1-3\eta^2 L^2 H^2)},$ we can get the $1/C'_1 \leq 2$. Thus,
\begin{flalign}
&\frac{1}{R}\sum_{r=0}^{R-1} \norm{ \triangledown {F} ( \bar{\boldsymbol{w}}^{r}) }_2^2 &\nonumber\\
&\leq \frac{2}{\eta R C'_1} (\mathbb{E} \left[F(\bar{\boldsymbol{w}}^{0}) \right] - F^{\star})+ \frac{\eta L(2+ C_1)\tau}{C_1'M} & \nonumber\\ 
&\quad + \frac{\sqrt{d}LG(1 + 2C_1) \sum_{i=1}^N \pi_i^2 \delta_{i}}{2C'_1} &\nonumber \\
%
& \leq \frac{4\Gamma}{\sqrt{MR}} +  \frac{6H L\tau}{\sqrt{MR}}+ \sqrt{d}LG \sum_{i=1}^N \pi_i^2 \delta_{i}, &
\end{flalign}
where $\Gamma = \mathbb{E} \left[F(\bar{\boldsymbol{w}}^{0}) \right] - F^{\star}$ and the proof is completed.




\section{Proof of Theorem~\ref{tm:oh}}\label{PF:oh}
\begin{proof}
It is obvious that variable $\epsilon_q$ is monotonously increasing with the function in (\ref{obj:hl}) with its feasible set, i.e., the optimal quantization error is $\epsilon_q = \epsilon_q^{\min}$. Substituting $\epsilon_q^{\star} = \epsilon_q^{\min}$ into problem (\ref{Eq:hl}) yields 
\begin{subequations} \label{Eq:hl_v1}
\begin{align}
\min_{H} & \quad \frac{(A_0H+A_1)^2}{MH(\epsilon-\epsilon_q^{\min})^2}(E^{cm}(\overline{\mathbf{B}})+H E^{cp}(\overline{\mathbf{q}}) ) \\
\text{s.t.} &  \quad   \rho(H) \leq MN(\epsilon-\epsilon_q^{\min})^2T_{\max}, \forall i, \label{c:hl}\\
& \quad H\geq 0, 
\end{align}
\end{subequations}
where $\rho(H) = (A_0H+A_1)^2 \left(\frac{T_i^{cm}({\overline{B}_{i}})}{H}  +T_i^{cp}(\overline{q}_i)\right)$. 
It is easy to verify that $\rho(H)$ is a convex function w.r.t $H$. Due to the convexity of $\rho(H)$, constraints (\ref{c:hl}) can be equivalently transformed to
\begin{flalign}
H_{\min} \leq H \leq H_{\max}, \label{c:hl_v1}
\end{flalign}
where $\rho(H_{\min}) = \rho(H_{\max}) = MN(\epsilon-\epsilon_q^{\min})^2T_{\max}$. By substituting (\ref{c:hl_v1}) into problem (\ref{Eq:hl}), we can simplify problem (\ref{Eq:hl_v1}) to (\ref{Eq:hl_r}).
\end{proof}

\section{Proof of Theorem~\ref{tm:oq}}\label{PF:oq}
\begin{proof}
For problem (\ref{Eq:qu1}), we observe that constraint (\ref{Constr:newq2}) is served as the upper bound of $\widetilde{q}_i^{(z)}$. Hence, we can substitute (\ref{Constr:newq2}) and (\ref{Constr:newq3}) constraints with
\begin{equation}
    1\leq \widetilde{q}_i^{(z)}\leq \widetilde{q}_i^{\max},\label{Constr:new_qr}
\end{equation}
where $\widetilde{q}_i^{\max} = C_i/U_i$. The Lagrangian duality $\mathcal{L}$ of (\ref{Eq:qu1}) is
\begin{flalign}
& \mathcal{L} (\mathbf{\widetilde{q}^{(z)}}, \mu^1, \boldsymbol{\mu}^2) \nonumber \\
&=R \sum_{i=1}^N p_i^{cp} (c_i^2 2^{\widetilde{q}_i^{(z)}}+ c_i^1)+ \mu^1 \left(\sum_{i=1}^N  \frac{A_3 \pi_i s}{2^{2^{\widetilde{q}_i^{(z)}}}-1} - \epsilon_q \right)  \nonumber\\
& + \sum_{i=1}^N \mu_{i}^2  \left( R(c_i^2 2^{\widetilde{q}_i^{(z)}}+ c_i^1) + \frac{K(H,\epsilon_q)\alpha_{i}^1}{B_{i}^{(z-1)}} - T_{\max} \right),   
\end{flalign}
where $\mu^1$ and $\boldsymbol{\mu}^2=\{\mu_{i}^2\}$ are the non-negative Lagrange multipliers corresponding to the related constraints. Since the objective function (\ref{obj:qu1}) is a monotonically decreasing function w.r.t $\widetilde{q}_i$, constraint~(\ref{Constr:newq1}) is always satisfied with equality. The Karush-Kuhn-Tucker~(KKT) conditions can be used to obtain the optimal solution, given by,
\begin{flalign}
 \left\{
\begin{array}{lr} 
\frac{\partial \mathcal{L}}{\partial \widetilde{q}_i^{(z)}} = 2^{\widetilde{q}_i^{(z)}}  \ln2 (R(p_i^{cp}+\mu_{i}^2)-  \ln2\frac{2^{2^{\widetilde{q}_i^{(z)}}}A_3 \pi_i^2 s \mu^1 }{ (2^{2^{\widetilde{q}_i^{(z)}}}-1 )^2}) = 0, &\\ 
\mu_{i}^2\left( R(c_i^2 2^{\widetilde{q}_i^{(z)}}+ c_i^1) + \frac{K(H,\epsilon_q)\alpha_{i}^1}{B_{i}^{(z-1)}} - T_{\max} \right) = 0, & \\ 
R(c_i^2 2^{\widetilde{q}_i^{(z)}}+ c_i^1) + \frac{K(H,\epsilon_q)\alpha_{i}^1}{B_{i}^{(z-1)}} - T_{\max} \leq 0,  \mu_{i}^2 \geq 0,\\
\sum_{i=1}^N  \frac{A_3 \pi_i^2 s}{2^{2^{\widetilde{q}_i^{(z)}}}-1} = \epsilon_q, \mu^1 \geq 0.&
\end{array} \label{KKT:QQ}
\right.
\end{flalign}
We can find $\widetilde{q}_i$ in closed-form as
\begin{flalign}
    %
    \widetilde{q}_i = \log_2\left(\log_2 (\lambda_i + \sqrt{\lambda_i^2+4}) - 1 \right), \label{KKT:q}
\end{flalign} 
where 
\begin{equation}
    \lambda_i = \frac{\ln(2)\mu^1A_3 \pi_i^2 s^2}{\frac{c_i^2(A_0H+A_1)^2}{MN (\epsilon-\epsilon_q)^2}(p_i^{cp}+\mu_{i}^2) }. \label{Eq:lambda}
\end{equation}
Here, $\widetilde{q}_i$ is a function of $\mu^{1}$ and $\mu_i^{2}$. We adopt the bisection search methods to find the optimal $\mu^{1\star}$ that satisfies constraint (\ref{Constr:newq1}). Given the KKT conditions, we then substitute the second equation in (\ref{KKT:QQ}) with (\ref{KKT:q}) and $\mu_i^{2\star}$ and calculate the optimal $\mu_i^{2\star}$.
 

For problem (\ref{Eq:b1}), constraint (\ref{Constr:newq3}) is equivalent to
\begin{flalign}
   B_{i}^{(z)} \geq B_{i,\min}^{(z)} \triangleq \frac{K(H,\epsilon_q)}{T_{\max} -RT^{cp}_i( 2^{\widetilde{q}_i^{(z)}}) }. \label{Constr:b1}
\end{flalign}
Substituting (\ref{Constr:b1}) into problem (\ref{Eq:b1}), we can obtain
\begin{subequations} 
\begin{align}
\underset{\mathbf{B}^{(z)}}{\min} & \quad K(H,\epsilon_q) \sum_{i=1}^N {p^{cm}_i\alpha_{i}^1}/{B_{i}^{(z)}} &\\
\text{s.t.} &\quad (\ref{Constr:Comm}),\\
&\quad B_{i}^{(z)} \geq B_{i,\min}^{(z)},&
\end{align}
\end{subequations}
and its Lagrangian is given as
\begin{flalign}
& \mathcal{L} (\mathbf{B}^{(z)}, \omega, \boldsymbol{\omega}^2)  \nonumber\\
&= \sum_{i=1}^N K(H,\epsilon_q)\frac{p^{cm}_i\alpha_{i}^1}{B_{i}^{(z)}} + \omega \left(\sum_{i=1}^N B_{i}^{(z)} - B_{\max}\right),
\end{flalign}
where $\omega \geq 0$ is the Lagrange multiplier associated with constraint~(\ref{Constr:Comm}). 
The KKT condition is then cast as
\begin{flalign}
\left\{
\begin{array}{lr} 
\frac{\partial \mathcal{L}}{\partial B_{i}^{(z)}} = - {p^{cm}K(H,\epsilon_q)\alpha_{i}^1}/{(B_{i}^{(z)})^2 } + \omega = 0,& \\
\sum_{i=1}^N B_{i}^{(z)} = B_{\max}.&
\end{array}\nonumber
\right. 
\end{flalign}
Therefore, we obtain the optimal $B_i^{\star}$ as in (\ref{Eq:opb}).
We then adopt the bisection search algorithm to efficiently find the optimal $\omega^{\star}$.
\end{proof}

\end{document}